\pdfoutput=1
\documentclass[acmsmall]{acmart}

\usepackage[english]{babel}

\usepackage{amsmath}
\usepackage{amsfonts}
\usepackage{bbm}
\usepackage{bm}
\usepackage{mathtools}
\usepackage{scalerel} %

\usepackage{booktabs} %
\usepackage{tabularx}
\usepackage{makecell}
\usepackage[online,flushleft]{threeparttable}
\usepackage{multirow}
\usepackage{subcaption}
\usepackage{adjustbox}
\usepackage{graphicx}
\usepackage[justification=justified, skip=5pt]{caption}

\usepackage[inline]{enumitem}
\usepackage[capitalize]{cleveref}

\usepackage{color}
\usepackage{tikz}
\usetikzlibrary{shapes, positioning, fit, calc}
\usetikzlibrary{arrows.meta}
\usetikzlibrary{shapes.arrows, shapes.geometric}
\usetikzlibrary{matrix}
\usepackage{tikzpeople}

\usepackage{pgfplots}
\usepgfplotslibrary{groupplots}
\pgfplotsset{compat=1.14}

\usepackage{epsfig}

\usepackage{soul}
\setstcolor{red}
\usepackage{sistyle}
\SIthousandsep{,}

\usepackage{algorithm}
\usepackage{algpseudocode}

\newcommand{\blue}[1]{\textcolor{black}{#1}}

\AtBeginDocument{%
  \providecommand\BibTeX{{%
    \normalfont B\kern-0.5em{\scshape i\kern-0.25em b}\kern-0.8em\TeX}}}

\definecolor{tea_green}{RGB}{214, 234, 193}
\definecolor{hint_green}{RGB}{226,246,209}
\definecolor{Madang}{RGB}{190,235,159}
\definecolor{yellow_green}{RGB}{198,222,119}
\definecolor{link_water}{RGB}{221, 232, 250}
\definecolor{celestial_blue}{RGB}{52, 152, 219}
\definecolor{shakespeare}{RGB}{85, 154, 193}
\definecolor{buttermilk}{RGB}{255,242,174}
\definecolor{chardonnay}{RGB}{250,196,114}
\definecolor{rajah}{RGB}{253,180,98}
\definecolor{fog}{RGB}{213, 193, 234}
\definecolor{melon}{RGB}{254,191,181}
\definecolor{sundown}{RGB}{249, 180, 181}
\definecolor{mona_lisa}{RGB}{246,152,134}
\definecolor{salmon}{RGB}{242,131,107}

\definecolor{saltpan}{RGB}{238, 243, 232}
\definecolor{aqua_spring}{RGB}{232, 243, 232}
\definecolor{tea_green}{RGB}{214, 234, 193}
\definecolor{Madang}{RGB}{190,235,159}
\definecolor{fringy_flower}{RGB}{194, 234, 193}
\definecolor{aero_blue}{RGB}{193, 234, 213}
\definecolor{pixie_green}{RGB}{183,214,170}
\definecolor{french_pass}{RGB}{195,232,246}
\definecolor{ice_cold}{RGB}{169,232,220}
\definecolor{pale_turquoise}{RGB}{172,240,242}
\definecolor{cruise}{RGB}{179,226,205}
\definecolor{sail}{RGB}{163,205,235}
\definecolor{spindle}{RGB}{179,205,227}
\definecolor{link_water}{RGB}{221, 232, 250}
\definecolor{periwinkle}{RGB}{203,213,232}
\definecolor{zanah}{RGB}{220, 233, 213}
\definecolor{frostee}{RGB}{217, 231, 214}
\definecolor{opal}{RGB}{199, 221, 211}
\definecolor{jet_stream}{RGB}{188, 214, 210}
\definecolor{skeptic}{RGB}{153, 187, 167}
\definecolor{hint_green}{RGB}{226,246,209}
\definecolor{snow_flurry}{RGB}{230,245,201}
\definecolor{surf_crest}{RGB}{205,230,208}
\definecolor{yellow_green}{RGB}{198,222,119}
\definecolor{cream}{RGB}{255,255,204}
\definecolor{pale_prim}{RGB}{255,255,179}
\definecolor{spring_sun}{RGB}{242,243,195}
\definecolor{portafino}{RGB}{245,237,160}
\definecolor{buttermilk}{RGB}{255,242,174}
\definecolor{cream_brulee}{RGB}{255, 229, 151}
\definecolor{dairy_cream}{RGB}{254,226,189}
\definecolor{champagne}{RGB}{254,217,166}
\definecolor{chardonnay}{RGB}{250,196,114}
\definecolor{manhattan}{RGB}{226,180,125}
\definecolor{rajah}{RGB}{253,180,98}
\definecolor{early_dawn}{RGB}{252,243,218}
\definecolor{egg_shell}{RGB}{238, 234, 215}
\definecolor{selago}{RGB}{243, 232, 243}
\definecolor{quartz}{RGB}{219,223,238}
\definecolor{fog}{RGB}{213, 193, 234}
\definecolor{languid_lavender}{RGB}{222,203,228}
\definecolor{watusi}{RGB}{254,221,207}
\definecolor{coral_andy}{RGB}{243,204,205}
\definecolor{cosmos}{RGB}{248,209,210}
\definecolor{melon}{RGB}{254,191,181}
\definecolor{azalea}{RGB}{234, 193, 194}
\definecolor{beauty_bush}{RGB}{235, 185, 179}
\definecolor{sundown}{RGB}{249, 180, 181}
\definecolor{mona_lisa}{RGB}{246,152,134}
\definecolor{salmon}{RGB}{242,131,107}

\definecolor{summer_sky}{RGB}{58, 151, 233}
\definecolor{chateau_green}{RGB}{72, 179, 96}
\definecolor{matisse}{RGB}{25, 104, 167}
\definecolor{allports}{RGB}{31, 106, 125}
\definecolor{sun_shade}{RGB}{255, 144, 68}
\definecolor{flamingo}{RGB}{237, 88, 85}
\definecolor{studio}{RGB}{128, 91, 160}

\definecolor{maya_blue}{RGB}{102, 204, 255}
\definecolor{feijoa}{RGB}{178,223,138}
\definecolor{sushi}{RGB}{117, 168, 47}
\definecolor{norway}{RGB}{158, 194, 132}
\definecolor{japanese_laurel}{RGB}{53, 116, 40}
\definecolor{see_green}{RGB}{161,228,195}
\definecolor{monte_carlo}{RGB}{135,204,194}
\definecolor{granny_smith_apple}{RGB}{150,214,150}
\definecolor{moss_green}{RGB}{170,216,176}
\definecolor{chateau_green}{RGB}{72, 179, 96}
\definecolor{opal}{RGB}{164,207,190}
\definecolor{acapulco}{RGB}{117, 170, 148}
\definecolor{viridian}{RGB}{55, 137, 122}
\definecolor{amazon}{RGB}{56, 123, 84}
\definecolor{asparagus}{RGB}{123, 160, 91}
\definecolor{fruit_salad}{RGB}{91, 160, 94}
\definecolor{puerto_rico}{RGB}{72, 179, 150}
\definecolor{mountain_meadow}{RGB}{0, 163, 136}
\definecolor{matisse}{RGB}{25, 104, 167}
\definecolor{allports}{RGB}{31, 106, 125}
\definecolor{astral}{RGB}{55, 111, 137}
\definecolor{spring_leaves}{RGB}{46, 83, 117}
\definecolor{biscay}{RGB}{44, 62, 80}
\definecolor{midnight}{RGB}{0, 29, 50}
\definecolor{amethyst}{RGB}{153, 102, 204}
\definecolor{studio}{RGB}{128, 91, 160}
\definecolor{tapestry}{RGB}{194, 109, 132}
\definecolor{atomic_tangerine}{RGB}{255, 153, 102}
\definecolor{amber}{RGB}{255, 191, 0}
\definecolor{casablanca}{RGB}{244, 178, 84}
\definecolor{california}{RGB}{233, 140, 58}
\definecolor{tomato}{RGB}{255, 97, 56} 
\definecolor{alizarin}{RGB}{233, 58, 64}

\definecolor{linen}{RGB}{251, 239, 227}
\definecolor{double_pearl_lusta}{RGB}{253, 242, 208}
\definecolor{oasis}{RGB}{253, 242, 208}
\definecolor{milan}{RGB}{255, 254, 169}
\definecolor{texas}{RGB}{245, 232, 123}
\definecolor{maize}{RGB}{249, 212, 156}

\definecolor{turmeric}{RGB}{211, 178, 76}
\definecolor{saffron}{RGB}{249,193,62}
\definecolor{my_sin}{RGB}{255, 176, 59}
\definecolor{tree_poppy}{RGB}{246, 154, 27}
\definecolor{jaffa}{RGB}{240, 131, 58}
\definecolor{crusta}{RGB}{254, 127, 44}
\definecolor{tahiti_gold}{RGB}{223, 102, 36}
\definecolor{outrageous_orange}{RGB}{255, 100, 45}
\definecolor{safety_orange}{RGB}{254, 106, 0}

\definecolor{azalea}{RGB}{251, 196, 196}
\definecolor{oyster_pink}{RGB}{238,206,205} 
\definecolor{coral_candy}{RGB}{242,208,205} 
\definecolor{baby_pink}{RGB}{246, 194, 192}
\definecolor{petite_orchid}{RGB}{223, 157, 155}
\definecolor{apricot}{RGB}{241,140,122}
\definecolor{NY_pink}{RGB}{228,136,113}
\definecolor{carmine_pink}{RGB}{231, 76, 60}
\definecolor{deep_carmine_pink}{RGB}{236, 50, 67}

\definecolor{wewak}{RGB}{244, 143, 150}
\definecolor{light_coral}{RGB}{244, 127, 123}
\definecolor{bittersweet}{RGB}{255,111,105}
\definecolor{carnation}{RGB}{245, 80, 86}
\definecolor{flamingo}{RGB}{237, 88, 85}
\definecolor{sunset_orange}{RGB}{242,89,75}
\definecolor{ku_crimson}{RGB}{243, 0, 25}
\definecolor{amaranth}{RGB}{234,46,73}
\definecolor{valencia}{RGB}{214, 87, 70}
\definecolor{chilean_fire}{RGB}{215, 87, 44}
\definecolor{mexican_red}{RGB}{170, 41, 37}

\definecolor{napa}{RGB}{163, 154, 137}

\definecolor{athens_gray}{RGB}{236, 240, 241}
\definecolor{gallery}{RGB}{240,240,240}
\definecolor{mercury}{RGB}{230,230,230}
\definecolor{platinum}{RGB}{228,228,228}
\definecolor{silver}{RGB}{191,191,191}
\definecolor{aluminum}{RGB}{153,153,153}
\definecolor{ship_gray}{RGB}{77,77,77}
\definecolor{tuatara}{RGB}{67, 67, 67}

\definecolor{malibu}{RGB}{110, 180, 240}
\definecolor{celestial_blue}{RGB}{52, 152, 219}
\definecolor{curious_blue}{RGB}{41, 128, 185}
\definecolor{french_blue}{RGB}{0, 112, 182}
\definecolor{matisse}{RGB}{25, 104, 167}
\definecolor{shakespeare}{RGB}{85, 154, 193}
\definecolor{seagull}{RGB}{128,177,211}
\definecolor{jelly_bean}{RGB}{45, 126, 150}
\definecolor{venice_blue}{RGB}{87, 135, 105}
\definecolor{boston_blue}{RGB}{68, 147, 161}

\definecolor{turquoise}{RGB}{41,217,194}
\definecolor{java}{RGB}{2,190,196}
\definecolor{riptide}{RGB}{141,211,199}
\definecolor{mountain_meadow}{RGB}{0, 163, 136}
\definecolor{free_speech_aquamarine}{RGB}{0, 156, 114}

\definecolor{cosmic_latte}{RGB}{222, 247, 229}
\definecolor{chinook}{RGB}{163, 232, 178}
\definecolor{padua}{RGB}{121, 189, 143}
\definecolor{ocean_green}{RGB}{79, 176, 112}
\definecolor{pastel_green}{RGB}{107, 227, 135}
\definecolor{chateau_green}{RGB}{69, 191, 85}
\definecolor{RoyalBlue}{RGB}{69, 191, 85}
\definecolor{pigment_green}{RGB}{0, 175, 79}
\definecolor{fern}{RGB}{101,197,117}
\definecolor{killarney}{RGB}{56, 113, 66}

\setcopyright{acmcopyright}
\copyrightyear{2022}
\acmYear{2022}
\acmDOI{10.1145/1122445.1122456}

\acmJournal{CSUR}
\acmVolume{37}
\acmNumber{4}
\acmArticle{111}
\acmMonth{4}

\begin{document}

\title{Graph Neural Networks in Recommender Systems: A Survey}

\author{Shiwen Wu}
\affiliation{%
  \department{School of CS and Key Lab of High Confidence Software Technologies (MOE)}
  \institution{Peking University}
  \city{Beijing}
  \postcode{100871}
  \country{China}
}

\email{wushw.18@pku.edu.cn}

\author{Fei Sun}
\authornote{Shiwen Wu and Fei Sun contributed equally to this research.}
\email{ofey.sf@alibaba-inc.com}
\affiliation{%
  \institution{Alibaba Group}
  \city{Beijing}
  \postcode{100102}
  \country{China}
}

\author{Wentao Zhang, Xu Xie, Bin Cui}
\authornote{Wentao Zhang and Bin Cui are the corresponding authors.}
\email{{wentao_zhang,xu.xie,bin.cui}@pku.edu.cn}
\affiliation{%
  \department{School of CS and Key Lab of High Confidence Software Technologies (MOE), Institute of Computational Social Science, Peking University (Qingdao)}
  \institution{Peking University}
  \city{Beijing}
  \postcode{100871}
  \country{China}
}

\renewcommand{\shortauthors}{Wu, et al.}

\begin{abstract}
With the explosive growth of online information, recommender systems play a key role to alleviate such information overload.
Due to the important application value of recommender systems, there have always been emerging works in this field.
\blue{In recommender systems, the main challenge is to learn the effective user/item representations from their interactions and side information (if any). 
Recently, graph neural network (GNN) techniques have been widely utilized in recommender systems since most of the information in recommender systems essentially has graph structure and GNN has superiority in graph representation learning.}
This article aims to provide a comprehensive review of recent research efforts on GNN-based recommender systems.
\blue{Specifically, we provide a taxonomy of GNN-based recommendation models according to the types of information used and recommendation tasks.
Moreover, we systematically analyze the challenges of applying GNN on different types of data and discuss how existing works in this field address these challenges.}
Furthermore, we state new perspectives pertaining to the development of this field.
\blue{We collect the representative papers along with their open-source implementations in \url{https://github.com/wusw14/GNN-in-RS}.}
\end{abstract}

\ccsdesc[500]{Information Systems~Recommender systems}

\keywords{Recommender System; Graph Neural Network; Survey}

\maketitle

\section{Introduction}
With the rapid development of e-commerce and social media platforms, recommender systems have become indispensable tools for many businesses~\cite{Covington:recsys16:Deep,ying2018graph,zhang2019deep,li2020lstm,chen2020co,xie2021causcf}.
They can be recognized as various forms depending on industries, like product suggestions on online e-commerce websites (e.g., Amazon and Taobao) or playlist generators for video and music services (e.g., YouTube, Netflix, and Spotify).
Users rely on recommender systems to alleviate the information overload problem and explore what they are interested in from the vast sea of items (e.g., products, movies, news, or restaurants).
Therefore, accurately modeling users' preferences from their historical interactions (e.g., click, watch, read, and purchase) lives at the heart of an effective recommender system.

Broadly speaking, in the past decades, the mainstream modeling paradigm in recommender systems has evolved from neighborhood methods~\cite{Herlocker:sigir99:Algorithmic,Sarwar:www01:Item,Linden:IC03:Amazon,Bell:icdm2007:Scalable} to representation learning based frameworks~\cite{Koren:2011:Advances,koren2009matrix,Wang:kdd15:Collaborative,Sedhain:www15:AutoRec,Covington:recsys16:Deep}.
Item-based neighborhood methods~\cite{Sarwar:www01:Item,Linden:IC03:Amazon,Bell:icdm2007:Scalable} directly recommend items to users that are similar to the historical items they have interacted with.
In a sense, they represent uses' preferences by directly using their historical interacted items.
Early item-based neighborhood approaches have achieved great success in real-world applications because of their simplicity, efficiency, and effectiveness.  %

An alternative approach is representation learning based methods that try to encode both users and items as continuous vectors (\emph{i.e.}, embeddings)
in a shared space, thus making them directly comparable.
Representation based models have sparked a surge of interest since the Netflix Prize competition~\cite{Bennett07thenetflix} demonstrates
matrix factorization models are superior to classic neighborhood methods for recommendations.
After that, various methods have been proposed to learn the representations of users and items 
from matrix factorization~\cite{koren2009matrix,Koren:2011:Advances} to deep learning models~\cite{Sedhain:www15:AutoRec,Covington:recsys16:Deep,he2017neural,zhang2019deep}.
Nowadays, deep learning models have been a dominant methodology for recommender systems in both academic research and industrial applications due to the ability in effectively capturing the non-linear and non-trivial user-item relationships and easily incorporating abundant data sources, e.g., contextual, textual, and visual information.

\blue{Among all those deep learning algorithms, one line is graph learning-based methods, which consider the information in recommender systems from the perspective of graphs~\cite{wang2020graph}.
Most of the data in recommender systems have a graph structure essentially~\cite{ying2018graph,van2018graph}.}
For example, the interaction data in a recommendation application can be represented by a bipartite graph between user and item nodes, with observed interactions represented by links.
Even the item transitions in users' behavior sequences can also be constructed as graphs.
The benefit of formulating recommendation as a task on graphs becomes especially evident when incorporating structured external information, e.g., the social relationship among users~\cite{fan2019graph,wu2019neural} and knowledge graph related to items~\cite{zhang2016collaborative,wang2019multi}.
\blue{In this way, graph learning provides a unified perspective to model the abundant heterogeneous data in recommender systems.
Early efforts in graph learning-based recommender systems utilize graph embedding techniques to model the relations between nodes, which can be further divided into factorization-based methods, distributed representation-based methods, and neural embedding-based methods~\cite{wang2020graph}.
Inspired by the superior ability of GNN in learning on graph-structured data, a great number of GNN-based recommendation models have emerged recently.}

Nevertheless, providing a unified framework to model the abundant data in recommendation applications is only part of the reason for the widespread adoption of GNN in recommender systems.
Another reason is that, different from traditional methods that only implicitly capture the collaborative signals (\emph{i.e.}, using user-item interactions as the supervised signals%
), GNN can naturally and explicitly encode the crucial collaborative signal (\emph{i.e.}, topological structure) to improve the user and item representations.
In fact, using collaborative signals to improve representation learning in recommender systems is not a new idea that originated from GNN~\cite{gori2007itemrank,koren2008factorization,kabbur2013fism,xie:icde2021:explore,Zhang:tois2020:Graph}.
Early efforts, such as SVD++~\cite{koren2008factorization} and FISM~\cite{kabbur2013fism}, have already demonstrated the effectiveness of the interacted items in user representation learning.
In view of the user-item interaction graph, these previous works can be seen as using one-hop neighbors to improve user representation learning.
The advantage of GNN is that it provides powerful and systematic tools to explore multi-hop relationships which have been proven to be beneficial to the recommender systems~\cite{ying2018graph,wang2019neural,he2020lightgcn}.

With these advantages, GNN has achieved remarkable success in recommender systems in the past few years.
In academic research, a lot of works demonstrate that GNN-based models outperform previous methods and achieve new state-of-the-art results on the public benchmark datasets~\cite{zhou2019data,wang2019neural,he2020lightgcn}.
Meanwhile, plenty of their variants are proposed and applied to various recommendation tasks, e.g., session-based recommendation~\cite{wu2019session,qiu2019rethinking}, Points-of-interest (POI) recommendation~\cite{chang2020learning,lim2020stp,wu2020garg}, group recommendation~\cite{he2020game,wang2020group}, multimedia recommendation~\cite{wei2019mmgcn,wei2020graph} and bundle recommendation~\cite{chang2020bundle}.
In industry, GNN has also been deployed in web-scale recommender systems to produce high-quality recommendation results~\cite{Eksombatchai:www17:Pixie,ying2018graph,Pfadler:icde20:Billion}.
For example, Pinterest developed and deployed a random-walk-based Graph Convolutional Network (GCN) algorithm model named PinSage on a graph with 3 billion nodes and 18 billion edges, and gained substantial improvements in user engagement in online A/B test.

\noindent\textbf{Differences between this survey and existing ones}.
\blue{There exist surveys focusing on different perspectives of recommender systems~\cite{quadrana2018sequence,zhang2019deep,batmaz2019review,guo2020survey,da2020recommendation,chen2020bias,chicaiza2021comprehensive}. However, there are very few comprehensive reviews that position existing works and current progress of applying GNN in recommender systems.
For example, \citet{zhang2019deep} and \citet{batmaz2019review} focus on most of the deep-learning techniques in recommender systems while ignoring GNN.
\citet{chen2020bias} summarizes the studies on the bias issue in recommender systems.
\citet{guo2020survey} review knowledge graph-based recommendations, and \citet{wang2021survey} propose a comprehensive survey in the session-based recommendations.
These two works only include some of the GNN methods applied in the corresponding sub-fields and examine a limited number of works.
To the extent of our knowledge, the most relevant survey published formally is a short paper~\cite{wang2020graph}, which presents a review of graph learning-based systems and briefly discusses the application of GNN in recommendation.
One recent survey under review~\cite{gao2021graph} classifies the existing works in GNN-based recommender systems from four perspectives of recommender systems, \emph{i.e.}, stage, scenario, objective, and application.
Such taxonomy emphasizes recommender systems but pays insufficient attention to applying GNN techniques in recommender systems.
Besides, this survey~\cite{gao2021graph} provides few discussions on the advantages and limitations of existing methods.
There are some comprehensive surveys on the GNN techniques~\cite{zhou2018graph,wu2020comprehensive}, but they only roughly discuss recommender systems as one of the applications.}

\blue{Given the impressive pace at which the GNN-based recommendation models are growing, we believe it is important to summarize and describe all the representative methods in one unified and comprehensible framework.
This survey summarizes the literature on the advances of GNN-based recommendation and discusses open issues or future directions in this field
To this end, more than 100 studies were shortlisted and classified in this survey.}

\noindent\textbf{Contribution of this survey}.
The goal of this survey is to thoroughly review the literature on the advances of GNN-based recommender systems and discuss further directions.
The researchers and practitioners who are interested in recommender systems could have a general understanding of the latest developments in the field of GNN-based recommendation.
The key contributions of this survey are summarized as follows:

$\bullet$ \textbf{New taxonomy}. We propose a systematic classification schema to organize the existing GNN-based recommendation models. 
\blue{Specifically, we categorize the existing works based on the type of information used and recommendation tasks into five categories: user-item collaborative filtering, sequential recommendation, social recommendation, knowledge graph-based recommendation, and other tasks (including POI recommendation, multimedia recommendation, etc.).}

$\bullet$ \textbf{Comprehensive review}. 
For each category, we demonstrate the main issues to deal with. Moreover, we introduce the representative models and illustrate how they address these issues.

$\bullet$ \textbf{Future research}. \blue{We discuss the limitations of current methods and propose nine potential future directions.}

\blue{The remaining of this article is organized as follows: Section 2 introduces the preliminaries for recommender systems and graph neural networks. Then, it discusses the motivations of applying GNN in recommender systems and categorizes the existing GNN-based recommendation models.
Section 3-7 summarizes the main issues of models in each category and how existing works tackle these challenges, and analyze their advantages and limitations.
Section 8 gives a summary of the mainstream benchmark datasets, widely-adopted evaluation metrics and real-world applications.}
Section 9 discusses the challenges and points out nine future directions in this field.
Finally, we conclude the survey in Section 10.

\section{Backgrounds and Categorization}
\label{sec:background}

Before diving into the details of this survey, we give a brief introduction to recommender systems and GNN techniques.
We also discuss the motivation of utilizing GNN techniques in recommender systems.
Furthermore, we propose a new taxonomy to classify the existing GNN-based models.
\blue{Throughout this paper, we use bold uppercase characters to denote matrices, bold lowercase characters to denote vectors, italic bold uppercase characters to denote sets, and calligraphic fonts to denote graphs.}
For easy reading, we summarize the notations that will be used throughout the paper in Table~\ref{tab:sym}.

\begin{table}
    \centering
    \caption{Key notations used in this paper}
    \resizebox{0.6\linewidth}{!} {
    \begin{tabular}{l|l}
    \toprule
    \textbf{Notations} & \textbf{Descriptions}\\
    \midrule
    $\mathcal{U}$/$\mathcal{I}$ & The set of users/items \\
    $\mathbf{R}=\{r_{u,i}\}$ & Interaction between users and items \\
    $\mathcal{G}_{\mathrm{S}}$ & Social relationship between users \\
    $\mathcal{G}_{\mathrm{KG}}$ & Knowledge graph \\
    $\mathcal{E}_{\mathrm{KG}}=\{e_i\}$ & The set of entities in Knowledge graph \\
    $\mathcal{R}_{\mathrm{KG}}=\{r_{e_i,e_j}\}$ & The set of relations in Knowledge graph \\
    $\mathbf{A}$ & Adjacency matrix of graph \\
    $\mathbf{A}^{\mathrm{in}}$/$\mathbf{A}^{\mathrm{out}}$ & In \& out adjacency matrix of directed graph \\
    $\mathcal{N}_v$ & Neighborhood set of node $v$\\
    $\mathbf{h}_v^{(l)}$ & Hidden state of node embedding at layer $l$\\
    $\mathbf{n}_v^{(l)}$ & Aggregated representation of node $v$'s neighbors at layer $l$\\
    $\mathbf{h}_u^{*}$ & Final representation of user $u$\\
    $\mathbf{h}_i^{*}$ & Final representation of item $i$\\
    $\mathbf{h}_u^{S}$ & Final representation of user $u$ in the social space\\
    $\mathbf{h}_u^{I}$ & Final representation of user $u$ in the item space\\
    $\mathbf{W}^{(l)}$ & Transformation matrix at layer $l$\\
    $\mathbf{W}_r^{(l)}$ & Transformation matrix of relation $r$ at layer $l$\\
    $\mathbf{b}^{(l)}$ & Bias term at layer $l$\\
    $\oplus$ & Vector concatenation \\
    $\odot$ & Element-wise multiplication operation \\
    \bottomrule
    \end{tabular}
    }
    \label{tab:sym}
\end{table}

\subsection{Recommender Systems}
Recommender systems infer users' preferences from user-item interactions or static features, and further recommend items that users might be interested in \cite{adomavicius2005toward}.
It has been a popular research area for decades because it has great application value and the challenges in this field are still not well addressed.
\blue{Formally, the task is to estimate her/his preference for any item $i\in \mathcal{I}$ by the learnt user representation $h_u^*$ and item representation $h_i^*$, i.e., 
\begin{equation}
    y_{u,i}=f(h_u^*,h_i^*),
\end{equation}
where score function $f(\cdot)$ can be dot product, cosine, multi-layer perceptions, etc., and $y_{u,i}$ denotes the preference score for user $u$ on item $i$, which is usually presented in probability.}

\blue{According to the types of information used to learn user/item representations, the research of recommender systems can usually be classified into specific types of tasks.
The \textit{user-item collaborative filtering recommendation} aims to capture the collaborative signal by leveraging only the user-item interactions, i.e., the user/item representations are jointly learned from pairwise data~\cite{lee2001algorithms, mnih2008probabilistic, koren2009matrix,Sedhain:www15:AutoRec,Wu:wsdm16:Collaborative,he2017neural}.
When the timestamps of the user's historical behavior are known or the historical behavior is organized in chronological order, the user representations can be enhanced via exploring the sequential patterns in her/his historical interactions~\cite{rendle2010factorizing,he2016fusing,tan2016improved,li2017neural,hidasi2018recurrent,liu2018stamp,kang2018self,sun:cikm2019:BERT4Rec,wang2021survey}.
According to whether the users are anonymous or not and whether the behaviors are segmented into sessions, works in this field can be further divided into \textit{sequential recommendation} and \textit{session-based recommendation}.
The session-based recommendation can be viewed as a sub-type of sequential recommendation with anonymous and session assumptions~\cite{quadrana2018sequence}.
In this survey, we do not distinguish them and refer to them collectively as the much broader term sequential recommendation for simplicity since our main focus is the contribution of GNN to recommendation, and the differences between them are negligible for the application of GNN.
In addition to sequential information, another line of research exploits the social relationship to enhance the user representations, which is classified as \textit{social recommendation}~\cite{ma2008sorec,jamali2010matrix,ma2011recommender,tang2013exploiting,ma2009learning,guo2015trustsvd}.
The social recommendation assumes that the users with social relationships tend to have similar user representations based on the social influence theory that connected people would influence each other.
Besides the user representation enhancement, a lot of efforts try to enhance the item representations by leveraging knowledge graph, which expresses relationships between items through attributes.
These works are always categorized as \textit{knowledge graph-based recommender systems}, which incorporates the semantic relations among items into collaborative signals.}

\subsection{Graph Neural Network Techniques}
Recently, systems based on variants of GNN have demonstrated ground-breaking performances on many tasks related to graph data, such as physical systems~\cite{battaglia2016interaction,sanchez2018graph}, protein structure~\cite{fout2017protein}, and knowledge graph~\cite{hamaguchi2017knowledge}.
\blue{In this part, we firstly introduce the definition of graphs, and then give a brief summary of the existing GNN techniques.}

\blue{A graph is represented as $\mathcal{G}=(\mathcal{V},\mathcal{E})$, where $\mathcal{V}$ is the set of nodes and $\mathcal{E}$ is the set of edges. Let $v_i\in \mathcal{V}$ be a node and $e_{ij}=(v_i, v_j)\in \mathcal{E}$ be an edge pointing from $v_j$ to $v_i$. The neighborhood of a node $v$ is denoted as $\mathcal{N}(v)=\{u\in\mathcal{V} | (v,u)\in \mathcal{E} \}$.
Generally, graphs can be categorized as:}

$\bullet$ \textbf{Directed/Undirected Graph}. A directed graph is a graph with all edges directed from one node to another. An undirected graph is considered as a special case of directed graphs where there is a pair of edges with inverse directions if two nodes are connected.

$\bullet$ \textbf{Homogeneous/Heterogeneous Graph}. Homogeneous graph consists of one type of nodes and edges, and heterogeneous graph has multiple types of nodes or edges.

$\bullet$ \textbf{Hypergraph}. A hypergraph is a generalization of a graph in which an edge can join any number of vertices. 

Given the graph data, the main idea of GNN is to iteratively aggregate feature information from neighbors and integrate the aggregated information with the current central node representation during the propagation process~\cite{zhou2018graph,wu2020comprehensive}.
From the perspective of network architecture, GNN stacks multiple propagation layers, which consist of the aggregation and update operations. 
\blue{The formulation of propagation is}
\blue{
\begin{equation}
\setlength{\abovedisplayskip}{1pt}
\setlength{\belowdisplayskip}{1pt}
    \begin{aligned}
    \text{Aggregation:}\quad \mathbf{n}_v^{(l)} &=\operatorname{Aggregator}_{l}\left(\left\{\mathbf{h}_{u}^{{l}}, \forall u \in \mathcal{N}_{v}\right\}\right), \\
    \text{Update:}\quad \mathbf{h}_v^{(l+1)} &=\operatorname{Updater}_{l}(\mathbf{h}_v^{(l)},\mathbf{n}_v^{(l)}),
    \end{aligned}
\end{equation}
where $\mathbf{h}_{u}^{(l)}$ denotes the representation of node $u$ at $l^{th}$ layer, and $\operatorname{Aggregator}_{l}$ and $\operatorname{Updater}_{l}$ represent the function of aggregation operation and update operation at $l^{th}$ layer, respectively.}
In the aggregation step, existing works either treat each neighbor equally with the mean-pooling operation~\cite{li2015gated,hamilton2017inductive}, or differentiate the importance of neighbors with the attention mechanism~\cite{velivckovic2017graph}.
In the update step, the representation of the central node and the aggregated neighborhood will be integrated into the updated representation of the central node.
In order to adapt to different scenarios, various strategies are proposed to better integrate the two representations, such as GRU mechanism~\cite{li2015gated}, concatenation with nonlinear transformation~\cite{hamilton2017inductive} and sum operation\cite{velivckovic2017graph}.
To learn more about GNN techniques, we refer the readers to the surveys~\cite{zhou2018graph,wu2020comprehensive}.

Here, we briefly summarize the aggregation and update operations of five typical GNN frameworks which are widely adopted in the field of recommendation.

    $\bullet$ \textbf{GCN}~\cite{kipf2017semi} approximates the first-order eigendecomposition of the graph Laplacian to iteratively aggregate information from neighbors. Concretely, it updates the embedding by 
    \blue{\begin{equation}
        \begin{aligned}
        \text{Aggregation:} \quad \mathbf{n}_{v}^{(l)} = \sum_{j \in \mathcal{N}_{v}} d_{vv}^{-\frac{1}{2}} \tilde{a_{vj}} d_{jj}^{-\frac{1}{2}}\mathbf{h}^{(l)}_j,\quad
        \text{Update:} \quad\mathbf{h}_{v}^{(l+1)} = \delta\left(\mathbf{W}^{(l)}\mathbf{n}_{v}^{(l)} \right),
        \end{aligned}
    \end{equation} 
    where $\delta(\cdot)$ is the nonlinear activation function, like ReLU, $\mathbf{W}^{(l)}$ is the learnable transformation matrix for layer $l$, $\tilde{a_{vj}}$ is the adjacency weight ($\tilde{a_{vv}}=1$) and $d_{jj}=\Sigma_k \tilde{a_{jk}}$.}
    
     $\bullet$ \textbf{GraphSAGE}~\cite{hamilton2017inductive} samples a fixed size of neighborhood for each node, proposes mean/sum/max-pooling aggregator and adopts concatenation operation for update,
    \blue{\begin{equation}
     \begin{aligned}
        \text{Aggregation:}\quad  \mathbf{n}_{v}^{(l)} &= \operatorname{Aggregator}_{l} \left(\left\{\mathbf{h}_{u}^{{l}}, \forall u \in \mathcal{N}_{v}\right\}\right), \\
        \text{Update:}\quad  \mathbf{h}_{v}^{(l+1)} &=\delta\left(\mathbf{W}^{(l)} \cdot\bigl[\mathbf{h}_{v}^{(l)} \oplus \mathbf{n}_{v}^{(l)}\bigr]\right),
    \end{aligned}
    \end{equation}
    where $\operatorname{Aggregator}_{l}$ denotes the aggregation function at $l^{th}$ layer,} $\delta(\cdot)$ is the nonlinear activation function, and $\mathbf{W}^{(l)}$ is the learnable transformation matrix.
    
    $\bullet$ \textbf{GAT}~\cite{velivckovic2017graph} %
    assumes that the influence of neighbors is neither identical nor pre-determined by the graph structure, thus it differentiates the contributions of neighbors by leveraging attention mechanism and updates the vector of each node by attending over its neighbors,
    \blue{\begin{equation}
        \begin{aligned}
            \text{Aggregation:} \quad &  
            \mathbf{n}_{v}^{(l)} =\sum_{j \in \mathcal{N}_{v}} \alpha_{v j}  \mathbf{h}^{(l)}_{j}, \alpha_{v j} =\frac{\exp\left(\text{Att}({h}^{(l)}_{v},\mathbf{h}^{(l)}_{j})\right)}{\sum_{k \in \mathcal{N}_{v}} \exp \left(\text{Att}({h}^{(l)}_{v},\mathbf{h}^{(l)}_{k})\right)},\\
            \text{Update:} \quad & \mathbf{h}_{v}^{(l+1)} = \delta\Bigl(\mathbf{W}^{(l)}\mathbf{n}_{v}^{(l)}\Bigr),
        \end{aligned}
    \end{equation}
    where $\text{Att}(\cdot)$ is a attention function and a typical $\text{Att}(\cdot)$ is $\operatorname{LeakyReLU}\left(\mathbf{a}^{T}\bigl[\mathbf{W}^{(l)}\mathbf{h}^{(l)}_{v} \oplus \mathbf{W}^{(l)}\mathbf{h}^{(l)}_{j}\bigr]\right)$,} $\mathbf{W}^{(l)}$ is responsible for transforming the node representations at $l^{th}$ propagation and $\mathbf{a}$ is the learnable parameter.
    
    $\bullet$ \textbf{GGNN}~\cite{li2015gated} adopts a gated recurrent unit (GRU)~\cite{li2015gated} in the update step,
    \blue{\begin{equation}
        \begin{aligned}
        \text{Aggregation:} \quad \mathbf{n}_v^{(l)} = \frac{1}{|\mathcal{N}_{v}|}\sum_{j \in \mathcal{N}_{v}}\mathbf{h}^{(l)}_j, \quad
        \text{Update:}\quad  \mathbf{h}_v^{(l+1)} = \text{GRU}(\mathbf{h}_v^{(l)},\mathbf{n}_v^{(l)}).
        \end{aligned}
    \end{equation}}
    \blue{GGNN executes the recurrent function several times over all nodes~\cite{wu2020comprehensive}, which might face the scalability issue when it is applied in large graphs.}
    
    $\bullet$ \blue{\textbf{HGNN}~\cite{feng2019hypergraph} is a typical hypergraph neural network, which encodes high-order data correlation in a hypergraph structure. The hyperedge convolutional layer is in the following formulation:
    \begin{equation}
    \begin{aligned}
        \text{Aggregation:}\quad \mathbf{N}^{(l)} = \mathbf{D}_v^{-\frac{1}{2}} \mathbf{E}\mathbf{W}^0\tilde{\mathbf{D_e}}^{-1}\mathbf{E}^T\mathbf{D}_v^{-\frac{1}{2}}\mathbf{H}^{(l)},\quad
        \text{Update:} \quad\mathbf{H}^{(l+1)} =\delta\left(\mathbf{W}^{(l)}\mathbf{N}^{(l)}\right),
    \end{aligned}
    \end{equation}
    where $\delta(\cdot)$ is the nonlinear activation function, like ReLU, $\mathbf{W}^{(l)}$ is the learnable transformation matrix for layer $l$, $\mathbf{E}$ is the hypergraph adjacent matrix, and $\mathbf{D}_e$ and $\mathbf{D}_v$ denote the diagonal matrices of the edge degrees and the vertex degrees, respectively.}

\subsection{Why Graph Neural Network for Recommendation}
In the past few years, many works on GNN-based recommendation have been proposed.
Before diving into the details of the latest developments, it is beneficial to understand the motivations of applying GNN to recommender systems.

The most intuitive reason is that GNN techniques have been demonstrated to be powerful in representation learning for graph data in various domains~\cite{zhou2018graph,guo2021syntax}, and most of the data in recommendation has essentially a graph structure as shown in Figure~\ref{fig:taxonomy}.
\blue{For instance, the user-item interaction data can be represented by a bipartite graph (as shown in Figure~\ref{fig:bipartite}) between the user and item nodes, where the link represents the interaction between the corresponding user and item.}
Besides, a sequence of items can be transformed into the sequence graph, where each item can be connected with one or more subsequent items. 
Figure~\ref{fig:seq} shows an example of sequence graph where there is an edge between consecutive items.
Compared to the original sequence data, sequence graph allows more flexibility to item-to-item relationships.
Beyond that, some side information also has naturally a graph structure, such as social relationship and knowledge graph, as shown in Figure~\ref{fig:social} and \ref{fig:kg}.

\tikzset{
  FARROW/.style={arrows={-{Latex[length=1.25mm, width=1.mm]}}, }, %
  U/.style = {circle, draw=melon!400, fill=melon, minimum width=1.4em, align=center, inner sep=0, outer sep=0},
  I/.style = {circle, draw=tea_green!400, fill=tea_green, minimum width=1.4em, align=center, inner sep=0, outer sep=0},
  cate/.style = {rectangle, draw, minimum width=2em, minimum height=1.6em, align=center, rounded corners=3, font=\scriptsize}, %
}

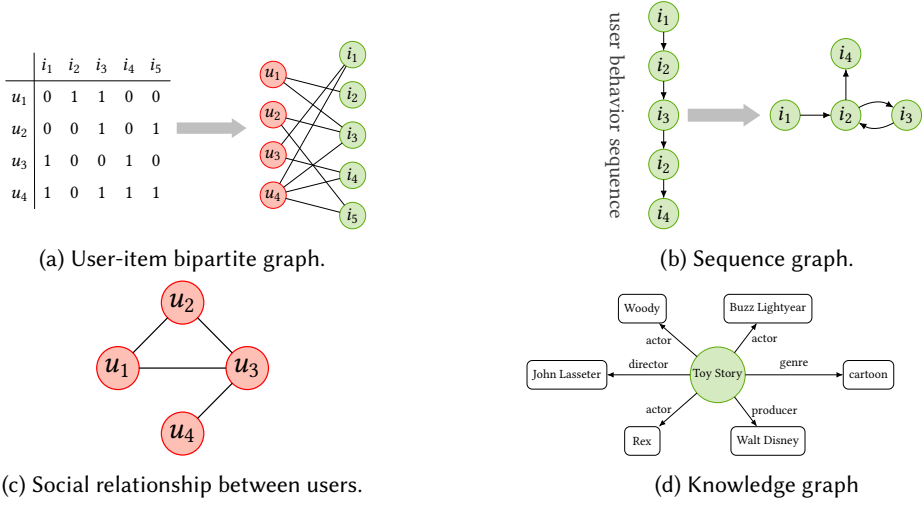
\begin{figure}
    \centering
    \begin{subfigure}[b]{0.45\textwidth}
    \centering
    \resizebox{0.8\linewidth}{!}{
        \begin{tikzpicture}[cell/.style={rectangle, text depth=0.5ex, text height=2ex},
space/.style={matrix of nodes, minimum height=1.5em, minimum width=1.5em, row sep=-\pgflinewidth, column sep=-\pgflinewidth }, nodes in empty cells]
            
    \matrix (m) [space, nodes={cell}, ampersand replacement=\&]
    {
     {} \& $i_1$ \& $i_2$ \& $i_3$ \& $i_4$ \& $i_5$ \\
    $u_1$  \& 0 \& 1 \& 1 \& 0 \& 0  \\  
    $u_2$  \& 0 \& 0 \& 1 \& 0 \& 1  \\ 
    $u_3$  \& 1 \& 0 \& 0 \& 1 \& 0  \\
    $u_4$  \& 1 \& 0 \& 1 \& 1 \& 1  \\   };
    \draw[black] (m-1-1.south west) -- (m-1-6.south east);
    \draw[black] ([yshift=-1mm] m-1-1.north east) -- (m-5-1.south east);
    
    \node [U, right of=m, node distance=3.5cm, yshift=0.25cm, align=center] (u2) {$u_2$};
    \node [U, above of=u2, node distance=0.75cm, align=center] (u1) {$u_1$};
    \node [U, below of=u2, node distance=0.75cm, align=center] (u3) {$u_3$};
    \node [U, below of=u3, node distance=0.75cm, align=center] (u4) {$u_4$};
    
     \foreach \x in {1,2,3,4}
    {
        \node [I, right of=u\x, node distance=1.5cm, yshift=0.37cm] (i\x) {$i_{\x}$};
    } 
    \node [I, below of=i4, node distance=0.75cm,] (i5) {$i_{5}$};
    \draw[] (u1) -- (i2);
    \draw[] (u1) -- (i3) -- (u2) -- (i5) -- (u4);
    \draw[] (u3) -- (i1) -- (u4) -- (i3);
    \draw (u4) -- (i4) -- (u3);
    
    \draw[-{Triangle[width=10pt, length=8pt]}, line width=6pt, silver] (m) -- ++(3, 0);
    
        \end{tikzpicture}
        }
        \caption{User-item bipartite graph.}
        \label{fig:bipartite}
    \end{subfigure}
\hfill
\begin{subfigure}[b]{0.45\textwidth}
\centering
\resizebox{0.7\linewidth}{!}{
        \begin{tikzpicture}[]
    
     \node [I, node distance=4cm, yshift=-0.5cm] (is1) at (0, 0) {$i_{1}$};
     \node [I, below of=is1, node distance=0.8cm] (is2)  {$i_{2}$};
     \node [I, below of=is2, node distance=0.8cm] (is3) {$i_{3}$};
     \node [I, below of=is3, node distance=0.8cm] (is4) {$i_{2}$};
     \node [I, below of=is4, node distance=0.8cm] (is5) {$i_{4}$};
     \node [align=center, left of=is3, node distance=0.8cm, rotate=-90, gray!150] (ranking) {user behavior sequence};
     
    \node [I, right of=is3, node distance=2cm] (i1) {$i_{1}$};
    \node [I, right of=i1, node distance=1cm,] (i2) {$i_2$};
    \node [I, right of=i2, node distance=1cm,] (i3) {$i_3$};
    \node [I, above of=i2, node distance=1cm,] (i4) {$i_4$};
    \draw[FARROW] (i1) -> (i2);
    \draw[FARROW] (i2) -> (i4);
    \draw[FARROW] (i2) edge[bend left=30] (i3);
    \draw[FARROW] (i3) edge[bend right=-30] (i2);
    \draw[FARROW] (is1) -> (is2);
    \draw[FARROW] (is2) -> (is3);
    \draw[FARROW] (is3) -> (is4);
    \draw[FARROW] (is4) -> (is5);
    
    \draw[-{Triangle[width=10pt, length=8pt]}, line width=6pt, silver] ($(is3.east)+(0.15, 0)$) -- ++(1.2, 0);
    
        \end{tikzpicture}}
        \caption{Sequence graph.}
        \label{fig:seq}
    \end{subfigure}

    \begin{subfigure}[b]{0.45\textwidth}
    \centering
    \resizebox{0.4\linewidth}{!}{
        \begin{tikzpicture}[]
            
    \node [U, node distance=3.5cm, align=center] (u1) at (0, 0){$u_1$};
    \node [U, above of=u1, node distance=0.75cm, xshift=0.75cm] (u2) {$u_2$};
    \node [U, right of=u1, node distance=1.50cm, align=center] (u3) {$u_3$};
    \node [U, below of=u1, node distance=0.75cm, xshift=0.75cm] (u4) {$u_4$};

    \draw[] (u1) -- (u2);
    \draw[] (u2) -- (u3) -- (u1);
    \draw[] (u3) -- (u4);
    
        \end{tikzpicture}
        }
        \caption{Social relationship between users.}
        \label{fig:social}
    \end{subfigure}
    \hfill
    \begin{subfigure}[b]{0.45\textwidth}
     \resizebox{0.8\linewidth}{!}{
        \centering
        \begin{tikzpicture}
            
    \node[cate] (o) {John Lasseter};
    \node [I, right of=o, node distance=3cm, align=center, yshift=0cm, inner sep=1] (movie) {\scriptsize Toy Story};
    \node [cate, right of=o, node distance=1.5cm, align=left, yshift=1.3cm] (actor1) {Woody};
    \node [cate, right of=o, node distance=4cm, align=left, yshift=1.3cm] (actor2) {Buzz Lightyear};
    \node [cate, right of=o, node distance=1.5cm, align=left, yshift=-1.3cm] (actor3) {Rex};
    \node [cate, right of=o, node distance=4cm, align=left, yshift=-1.3cm] (producer) {Walt Disney};
    \node [cate, right of=movie, node distance=3cm, align=left, yshift=0cm] (genre) {cartoon};
    \node [above of=actor1, node distance=.5cm] (ph) {};
    
    \draw[FARROW] (movie) -> (o) node[pos=0.5, above] {\scriptsize director};
    \draw[FARROW] (movie) -> (genre) node[pos=0.5, above] {\scriptsize genre} ;
    \draw[FARROW] (movie) -> (actor1) node[pos=0.5, left] {\scriptsize actor} ;
    \draw[FARROW] (movie) -> (actor2) node[pos=0.5, right] {\scriptsize actor} ;
    \draw[FARROW] (movie) -> (actor3) node[pos=0.5, left] {\scriptsize actor} ;
    \draw[FARROW] (movie) -> (producer) node[pos=0.5, right] {\scriptsize producer} ;
    \end{tikzpicture}
    }
    \caption{Knowledge graph}
    \label{fig:kg}
    \end{subfigure}
    \caption{Representative graph structures in recommender systems.}
    \label{fig:taxonomy}
\end{figure}
Due to the specific characteristic of different types of data in recommendation, a variety of models have been proposed to effectively learn their pattern for better recommendation results, which is a big challenge for the model design.
Considering the information in recommendation from the perspective of the graph, a unified GNN framework can be utilized to address all these tasks.
For example, the task of non-sequential recommendation is to learn the effective node representations, i.e., user/item representations, and to further predict user preferences.
The task of sequential recommendation is to learn the informative graph representation, i.e., sequence representation.
Both node representation and graph representation can be learned through GNN.
Besides, it is more convenient and flexible to incorporate additional information (if available), compared to non-graph perspective.
\blue{For instance, the social network can be integrated into the user-item bipartite relationship as a unified graph. Both the social influence and collaborative signal can be captured during the iterative propagation.}

Moreover, GNN can explicitly encode the crucial collaborative signal of user-item interactions to enhance the user/item representations through the propagation process.
Utilizing collaborative signals for better representation learning is not a completely new idea.
For instance, SVD++~\cite{koren2008factorization} incorporates the representations of interacted items to enrich the user representations.
ItemRank~\cite{gori2007itemrank} constructs the item-item graph from interactions, and adopts the random-walk algorithm to rank items according to user preferences.
Note that SVD++ can be seen as using one-hop neighbors (i.e., items) to improve user representations, while ItemRank utilizes two-hop neighbors to improve item representations.
Compared with non-graph model, GNN is more flexible and convenient to model multi-hop connectivity from user-item interactions, and the captured CF signals in high-hop neighbors have been demonstrated to be effective for recommendation.

\subsection{Categories of Graph Neural Network-Based Recommendation}
In this survey, we propose a new taxonomy to classify the existing GNN-based models.
\blue{Based on the types of information used and recommendation tasks, the existing works are categorized into user-item collaborative filtering, sequential recommendation, social recommendation, knowledge graph-based recommendation, and other tasks.
In addition to the former four types of tasks, there are other recommendation tasks, such as POI recommendation, multimedia recommendation, and bundle recommendation.
Since the studies utilizing GNN in these tasks are not that abundant, we group them into one category and discuss their current developments, respectively.}

\blue{The rationale of classification is as follows:
The graph structure depends to a large extent on the type of information.
For example, a social network is naturally a homogeneous graph, and user-item interaction can be considered either a bipartite graph or two homogeneous graphs (i.e., user-user and item-item graphs).
Besides, the information type also plays a key role in designing an efficient GNN architecture, such as aggregation and update operations and network depth.
For instance, a knowledge graph has multi-type entities and relations, which requires considering such heterogeneity during propagation.
Moreover, recommendation tasks are highly related to the type of information used.
For example, the social recommendation is to make a recommendation by utilizing the social network information, and the knowledge graph-based recommendation is to enhance the item representation by leveraging semantic relations among items in the knowledge graph.
This survey is mainly for the readers interested in the development of GNN in recommender systems. Thus our taxonomy is primarily from the perspective of recommender systems but also takes the GNN into account.}

\section{User-item Collaborative Filtering}
\blue{Given the user-item interaction data, the basic idea of user-item collaborative filtering is essentially using the items interacted by users to enhance user representations and using the users once interacted with items to enrich item representations.}
Inspired by the advantage of GNN techniques in simulating the information diffusion process, recent efforts have studied the design of GNN methods, in order to exploit high-order connectivity from user-item interactions more efficiently.
\blue{Figure~\ref{fig:user_item} illustrates the pipeline of applying GNN to user-item interaction information.}

To take full advantage of GNN methods on capturing collaborative signals from user-item interactions, there are four main issues to deal with:
\tikzset{
  FARROW/.style={arrows={-{Latex[length=1.25mm, width=1.mm]}}, }, %
  U/.style = {circle, draw=melon!400, fill=melon, minimum width=1.4em, align=center, inner sep=0, outer sep=0},
  I/.style = {circle, draw=tea_green!400, fill=tea_green, minimum width=1.4em, align=center, inner sep=0, outer sep=0},
  cate/.style = {rectangle, draw, minimum width=8em, minimum height=2em, align=center, rounded corners=3}, %
  cate2/.style = {rectangle, minimum width=2em, minimum height=2em, align=center, rounded corners=3},%
  encoder/.style = {rectangle, fill=Madang!82, minimum width=10em, minimum height=3em, align=center, rounded corners=3},
  v_rep/.style={
       rectangle split,
       rectangle split part align=base,
       rectangle split horizontal=true,
       rectangle split draw splits=true,
       rectangle split parts=5,
       rectangle split part fill={red!30, blue!20, athens_gray!80, matisse, silver},
       draw=gray, %
       very thin,
       minimum height=1em,
       minimum width=2em,
       inner sep=2.5pt,
       text centered,
       text=gray,
       rounded corners=1
       },
}

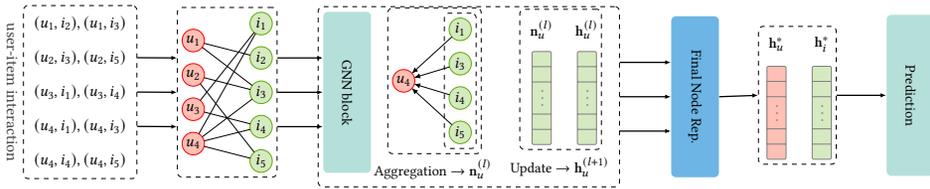
\begin{figure}
\centering
\resizebox{0.9\linewidth}{!}{
    \begin{tikzpicture}
     \node [node distance=4cm, yshift=-0.5cm] (is1) at (0, 0) {$(u_1,i_2),(u_1,i_3)$};
     \node [below of=is1, node distance=0.75cm] (is2)  {$(u_2,i_3),(u_2,i_5)$};
     \node [below of=is2, node distance=0.75cm] (is3)  {$(u_3,i_1),(u_3,i_4)$};
     \node [below of=is3, node distance=0.75cm] (is4)  {$(u_4,i_1),(u_4,i_3)$};
     \node [below of=is4, node distance=0.75cm] (is5)  {$(u_4,i_4),(u_4,i_5)$};
     \node [align=center, left of=is3, node distance=1.5cm, rotate=-90, gray!150] (ranking) {user-item interaction};
     \node [dashed, draw=black, fit={(is1) (is5)}, inner sep=3, rounded corners=3] (sf) {};
     
    \node [U, right of=is3, node distance=2.5cm, yshift=0.375cm, align=center] (u2) {$u_2$};
    \node [U, above of=u2, node distance=0.75cm, align=center] (u1) {$u_1$};
    \node [U, below of=u2, node distance=0.75cm, align=center] (u3) {$u_3$};
    \node [U, below of=u3, node distance=0.75cm, align=center] (u4) {$u_4$};
    
     \foreach \x in {1,2,3,4}
    {
        \node [I, right of=u\x, node distance=1.5cm, yshift=0.37cm] (i\x) {$i_{\x}$};
    } 
    \node [I, below of=i4, node distance=0.75cm,] (i5) {$i_{5}$};
    \draw[] (u1) -- (i2);
    \draw[] (u1) -- (i3) -- (u2) -- (i5) -- (u4);
    \draw[] (u3) -- (i1) -- (u4) -- (i3);
    \draw (u4) -- (i4) -- (u3);
    \node [dashed, draw=black, fit={(u1)(i1)(i5)}, inner sep=3, rounded corners=3] (gf) {};
    
    \node[encoder, fill=monte_carlo!62, right of=i3, node distance=1.9cm, align=center, rotate=-90] (sg) {\small GNN block};
    
    \node [U, right of=sg, node distance=1.25cm, yshift=0.25cm] (u41) {$u_4$};
    \node [I, right of=u41, node distance=1.25cm, yshift=0.375cm] (i31) {$i_{3}$};
    \node [I, above of=i31, node distance=0.75cm] (i11) {$i_{1}$};
    \node [I, below of=i31, node distance=0.75cm] (i41) {$i_{4}$};
    \node [I, below of=i41, node distance=0.75cm] (i51) {$i_{5}$};
    \draw[FARROW] (i11) -- (u41);
    \draw[FARROW] (i31) -- (u41);
    \draw[FARROW] (i41) -- (u41);
    \draw[FARROW] (i51) -- (u41);
    \node [dashed, draw=black, fit={(i11)(i51)}, inner sep=3, rounded corners=3] (iagg) {};
    \node [dashed, draw=black, fit={(u41)(iagg)}, inner sep=3, rounded corners=3] (agg) {};
    \node [below of=i51, node distance=0.8cm, xshift=-0.6cm] (aggnote) {\small Aggregation $\rightarrow \mathbf{n}_u^{(l)}$};
    
    \node[v_rep, minimum height=1.2em, rectangle split part fill={tea_green}, right of=i41, node distance=1.8cm, rotate=-90] (nu) {\nodepart{three} $\,\cdots$};
    \node[v_rep, minimum height=1.2em, rectangle split part fill={tea_green}, right of=i41, node distance=2.8cm, rotate=-90] (hu) {\nodepart{three} $\,\cdots$};
    \node [above of=nu, node distance=1.5cm] (s1) {\small $\mathbf{n}_u^{(l)}$};
    \node [above of=hu, node distance=1.5cm] (s2) {\small $\mathbf{h}_u^{(l)}$};
    \node [dashed, draw=black, fit={(s1)(hu)}, inner sep=3, rounded corners=3] (up) {};
    \node [below of=hu, node distance=1.5cm, xshift=-0.6cm] (upnote) {\small Update $\rightarrow \mathbf{h}_u^{(l+1)}$};
    
    \node [dashed, draw=black, fit={(sg)(upnote)}, inner sep=3, rounded corners=3] (gnn) {};
    
    \node[encoder, fill=celestial_blue!72, right of=gnn, node distance=5cm, align=center, rotate=-90] (seq) {\small Final Node Rep.};
    
    \node[v_rep, minimum height=1.2em, rectangle split part fill={melon}, right of=seq, node distance=1.8cm, yshift=-0.35cm, rotate=-90] (us) {\nodepart{three} $\,\cdots$};
    \node [above of=us, node distance=1.5cm] (us_v) {\small $\mathbf{h}_u^*$};
    \node[v_rep, minimum height=1.2em, rectangle split part fill={tea_green}, right of=seq, node distance=2.8cm, yshift=-0.35cm, rotate=-90] (is) {\nodepart{three} $\,\cdots$};
    \node [above of=is, node distance=1.5cm] (is_v) {\small $\mathbf{h}_i^*$};
    \node [dashed, draw=black, fit={(us_v)(is)}, inner sep=3, rounded corners=3] (final) {};
    
    \node[encoder, fill=monte_carlo!62, right of=final, node distance=2.5cm, align=center, rotate=-90] (pred) {\small Prediction};
    
    \draw[FARROW, thick] (gnn.east) -- (seq.south);
    \draw[FARROW, thick] ($(gnn.east)+(0, -0.75)$) -- ($(seq.south)+(0, -0.75)$);
    \draw[FARROW, thick] ($(gnn.east)+(0, 0.75)$) -- ($(seq.south)+(0, 0.75)$);
    \draw[FARROW, thick] (seq.north) -- (final.west);
    \draw[FARROW, thick] (final.east) -- (pred.south);

      \draw[FARROW, thick] (sf.east) -- (gf.west);
      \draw[FARROW, thick] ($(sf.east)+(0, -0.75)$) -- ($(gf.west)+(0, -0.75)$);
      \draw[FARROW, thick] ($(sf.east)+(0, 0.75)$) -- ($(gf.west)+(0, 0.75)$);
      \draw[FARROW, thick] (gf.east) -- (sg.south);
      \draw[FARROW, thick] ($(gf.east)+(0, -0.75)$) -- ($(sg.south)+(0, -0.75)$);
      \draw[FARROW, thick] ($(gf.east)+(0, 0.75)$) -- ($(sg.south)+(0, 0.75)$);
      
    \end{tikzpicture}
    }
    \caption{\blue{The overall framework of GNN in user-item collaborative filtering.}}
    \label{fig:user_item}
\end{figure}

$\bullet$ \textbf{Graph Construction}. 
    Graph structure is essential for the scope and type of information to propagate.
    The original bipartite graph consists of a set of user/item nodes and the interactions between them. 
    Whether to apply GNN over the heterogeneous bipartite graph or construct the homogeneous graph based on two-hop neighbors?
    Considering computational efficiency, how to sample representative neighbors for graph propagation instead of operating on the full graph?
    
$\bullet$  \textbf{Neighbor Aggregation}. 
    How to aggregate the information from neighbor nodes? Specifically, whether to differentiate the importance of neighbors, model the affinity between the central node and neighbors, or the interactions among neighbors?

$\bullet$  \textbf{Information Update}. 
    How to integrate the central node representation and the aggregated representation of its neighbors? 

$\bullet$  \textbf{Final Node Representation}. 
    Predicting the user's preference for the items requires the overall user/item representation.
    Whether to use the node representation in the last layer or the combination of the node representations in all layers as the final node representation?

\subsection{Graph construction}
Most of works~\cite{van2018graph,zheng2018spectral,zhang2019star,wang2019neural,li2019hierarchical,chen2020revisiting,he2020lightgcn,wang2019binarized,tan2020learning,sun2020neighbor,wu2020joint} apply the GNN on the original user-item bipartite graph directly.
There are two issues in directly applying GNN on the original graph:
one is effectiveness that the original graph structure might not be sufficient enough for learning user/item representations;
another one is efficiency that aggregating the information of the full neighborhoods of nodes requires high computation cost especially for the large-scale graph~\cite{ying2018graph}.

\blue{One strategy to address the first issue is to enrich the original graph structure by adding edges, such as links between two-hop neighbors and hyperedges.}
For instance, Multi-GCCF~\cite{sun2019multi} and DGCF~\cite{liu2020deoscillated} add edges between two-hop neighbors on the original graph to obtain the user-user and item-item graph.
In this way, the proximity information among users and items can be explicitly incorporated into user-item interactions.
\blue{DHCF~\cite{ji2020dual} introduces the hyperedges and constructs the user/item hypergraphs, in order to capture explicit hybrid high-order correlations.
Another strategy is to introduce virtual nodes for enriching the user-item interactions.
For example, DGCF~\cite{wang2020disentangled} introduces virtual intent nodes and decomposes the original graph into a corresponding subgraph for each intent, which represents the node from different aspects and has better expressive power.
HiGNN~\cite{li2020hierarchical2} creates new coarsened user-item graphs by clustering similar users/items and taking the clustered centers as new nodes, in order to explicitly capture hierarchical relationships among users and items.}

In terms of the second issue, sampling strategies are proposed to make GNN efficient and scalable to large-scale graph-based recommendation tasks.
PinSage~\cite{ying2018graph} designs a random-walk based sampling method to obtain the fixed size of neighborhoods with the highest visit counts.
In this way, those nodes that are not directly adjacent to the central node may also become its neighbors.
Multi-GCCF~\cite{sun2019multi} and NIA-GCN~\cite{sun2020neighbor} randomly sample a fixed size of neighbors.
Sampling is a trade-off between the original graph information and computational efficiency.
The performance of the model depends on the sampling strategy and the more efficient sampling strategy for neighborhood construction deserves further studying.

\subsection{Neighbor aggregation}

The aggregation step is of the vital importance of information propagation for the graph structure, which decides how much neighbors' information should be propagated.
Mean-pooling is one of the most straightforward aggregation operations~\cite{van2018graph,zhang2019star,sun2019multi,zhang2019inductive,tan2020learning}, which treats neighbors equally, 
\begin{equation}
    \textbf{n}_u^{(l)}=\frac{1}{\left|\mathcal{N}_{u}\right|}\mathbf{W}^{(l)} \mathbf{h}_{i}^{(l)}.
    \label{eq:mean-pooling}
\end{equation}
Mean-pooling is easy for implementation but might be inappropriate when the importance of neighbors is significantly different.
Following the traditional GCN, some works employ ``degree normalization''~\cite{chen2020revisiting,he2020lightgcn,wu2020joint}, which assigns weights to nodes based on the graph structure,
\begin{equation}
    \textbf{n}_u^{(l)} =\sum_{i \in \mathcal{N}_{u}} \frac{1}{\sqrt{\left|\mathcal{N}_{u}\right|\left|\mathcal{N}_{i}\right|}}\mathbf{W}^{(l)} \mathbf{h}_{i}^{(l)}.
\end{equation}
Owing to the random-walk sampling strategy, PinSage~\cite{ying2018graph} adopts the normalized visit counts as the importance of neighbors when aggregating the vector representations of neighbors.
However, these aggregation functions determine the importance of neighbors according to the graph structure but ignore the relationships between the connected nodes.

\blue{Motivated by common sense that the embeddings of items in line with the user's interests should be passed more to the user (analogously for the items), MCCF~\cite{wang2019multic} and DisenHAN~\cite{ma2020disentangled} leverage attention mechanism to learn the weights of neighbors~\cite{wang2019multi, ma2020disentangled}.}
NGCF~\cite{wang2019neural} employs element-wise product to augment the items’ features the user cares about or the users’ preferences for features the item has.
Take the user node as an example, the aggregated neighbor representation is calculated as follows:
\begin{equation}
\textbf{n}_u^{(l)} =\sum_{i \in \mathcal{N}_{u}} \frac{1}{\sqrt{\left|\mathcal{N}_{u}\right|\left|\mathcal{N}_{i}\right|}}\left(\mathbf{W}_{1}^{(l)} \mathbf{h}_{i}^{(l)}+\mathbf{W}_{2}^{(l)}\Bigl(\mathbf{h}_{i}^{(l)} \odot \mathbf{h}_{u}^{(l)}\Bigr)\right).
\end{equation}
NIA-GCN~\cite{sun2020neighbor} argues that existing aggregation functions fail to preserve the relational information within the neighborhood, thus proposes the pairwise neighborhood aggregation approach to explicitly capture the interactions among neighbors.
Concretely, it applies element-wise multiplication between every two neighbors to model the user-user/item-item relationships.

\subsection{Information update}
\blue{Given the information aggregated from its neighbors, how to update the representation of the node is essential for iterative information propagation.
According to whether to retain the information of the node itself, the existing methods can be divided into two directions.
One is to discard the original information of the user or item node completely and use the aggregated representation of neighbors as the new central node representation~\cite{van2018graph,zhang2019star,he2020lightgcn,wang2020disentangled}, which might overlook the intrinsic user preference or the intrinsic item property.}

Another is to take both the node itself ($\mathbf{h}_u^{(l)}$) and its neighborhood message ($\mathbf{n}_u^{(l)}$) into consideration to update node representations.
The most straightforward way is to combine these two representations linearly with sum-pooling or mean-pooling operation~\cite{wang2019neural,zhang2019inductive,sun2020neighbor,wu2020joint}.
Inspired by the GraphSAGE~\cite{hamilton2017inductive}, some works~\cite{ying2018graph,li2019hierarchical,sun2019multi} adopt concatenation function with nonlinear transformation to integrate these two representations as follows:
\begin{equation}
\textbf{h}_u^{(l+1)} = \sigma\left(\mathbf{W}^{(l)} \cdot (\mathbf{h}_{u}^{(l)}\oplus \mathbf{n}_{u}^{(l)})+\mathbf{b}^{(l)}\right),
\end{equation}
where $\sigma$ denotes the activation function, e.g., ReLU, LeakyReLU, and sigmoid.
Compared to linear combination, concatenation operation with feature transformation allows more complex feature interaction.
LightGCN~\cite{he2020lightgcn} and LR-GCCF~\cite{chen2020revisiting} observe that nonlinear activation contributes little to the overall performance, and they simplify the update operation by removing the non-linearities, thereby retaining or even improving performance and increasing computational efficiency.

\subsection{Final node representation}
Applying the aggregation and update operations layer by layer generates the representations of nodes for each depth of GNN.
The overall representations of users and items are required for the final prediction task.

A mainstream approach is to use the node vector in the last layer as the final representation, \emph{i.e.}, $\textbf{h}_u^*=\textbf{h}_u^{(L)}$~\cite{van2018graph,ying2018graph,zhang2019star,li2019hierarchical,tan2020learning,wang2020disenhan}.
However, the representations obtained in different layers emphasize the messages passed over different connections~\cite{wang2019neural}.
Specifically, the representations in the lower layer reflect the individual feature more while those in the higher layer reflect the neighbor feature more.
To take advantage of the connections expressed by the output of different layers, recent studies employ different methods to integrate the messages from different layers.
\blue{\begin{equation}
    \begin{aligned}
        \text{Mean-pooling:} \quad  &\mathbf{h}_u^*=\frac{1}{L+1}\sum_{l=0}^{L}\mathbf{h}_u^{(l)},\\
        \text{Sum-pooling:}  \quad &\mathbf{h}_u^*=\sum_{l=0}^{L}\mathbf{h}_u^{(l)},\\
        \text{Weighted-pooling:} \quad &\mathbf{h}_u^*=\frac{1}{L+1}\sum_{l=0}^{L}\alpha^{(l)}\mathbf{h}_u^{(l)},\\
        \text{Concatenation:} \quad &\mathbf{h}_u^*=\mathbf{h}_u^{(0)}\oplus \mathbf{h}_u^{(1)}\oplus\cdots\oplus \mathbf{h}_u^{(L)},
    \end{aligned}
\end{equation}}
where $\alpha^{(l)}$ is a learnable parameter.
Note that mean-pooling and sum-pooling can be seen as two special cases of weighted pooling.
Compared to mean-pooling and sum-pooling, weighted pooling allows more flexibility to differentiate the contribution of different layers.
Among these four methods, the former three all belong to linear operation, and only concatenation operation preserves information from all layers.

\tikzset{
  FARROW/.style={arrows={-{Latex[length=1.25mm, width=1.mm]}}, }, %
  U/.style = {circle, draw=melon!400, fill=melon, minimum width=1.4em, align=center, inner sep=0, outer sep=0},
  I/.style = {circle, draw=tea_green!400, fill=tea_green, minimum width=1.4em, align=center, inner sep=0, outer sep=0},
  cate/.style = {rectangle, draw, minimum width=8em, minimum height=2em, align=left, rounded corners=3}, %
  cate2/.style = {rectangle, minimum width=2em, minimum height=2em, align=center, rounded corners=3},%
  encoder/.style = {rectangle, fill=Madang!82, minimum width=10em, minimum height=3em, align=center, rounded corners=3},
}

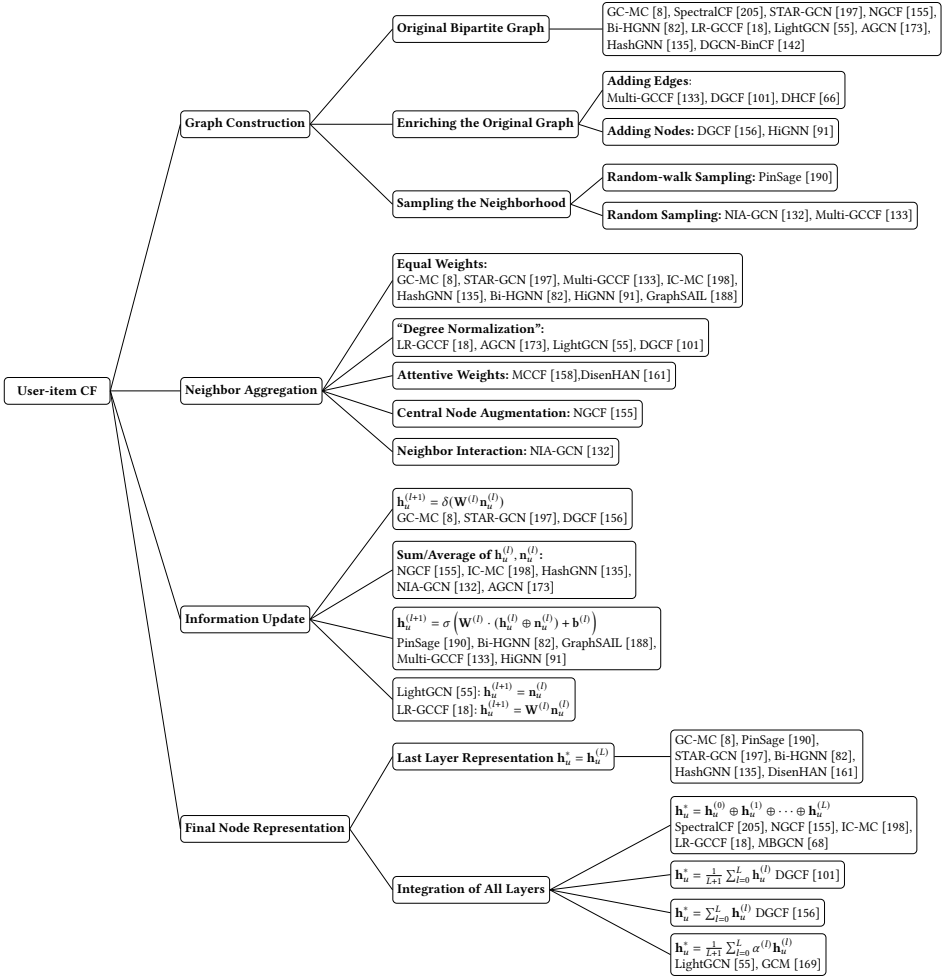
\begin{figure}
    \centering
    \resizebox{0.9\linewidth}{!}{
    \begin{tikzpicture}
    
    \node [cate, distance=4cm, yshift=-0.5cm, align=left] (n1) at (0, 0) {\textbf{User-item CF}};
    
    \node [cate, right of=n1, node distance=5cm, yshift=7cm,] (n21) {\textbf{Graph Construction}};
    \node [cate, below=7cm of n21.west, anchor=west] (n22) {\textbf{Neighbor Aggregation}};
    \node [cate, below=6cm of n22.west, node distance=5cm, anchor=west,] (n23) {\textbf{Information Update}};
    \node [cate, below=5.5cm of n23.west, anchor=west] (n24) {\textbf{Final Node Representation}};

    \draw[] (n1.east) -- (n21.west);
    \draw[] (n1.east) -- (n22.west);
    \draw[] (n1.east) -- (n23.west);
    \draw[] (n1.east) -- (n24.west);
    
    \node [cate, right of=n21, node distance=6cm, yshift=2.5cm,] (n211) {\textbf{Original Bipartite Graph}};
    \node [cate,  below=2.5cm of n211.west, anchor=west] (n212) {\textbf{Enriching the Original Graph}};
    \node [cate, below=2.1cm of n212.west, anchor=west] (n213) {\textbf{Sampling the Neighborhood}};
    
    \draw[] (n21.east) -- (n211.west);
    \draw[] (n21.east) -- (n212.west);
    \draw[] (n21.east) -- (n213.west);
    
    \node [cate, right of=n211, node distance=8cm, align=left] (n2111) {GC-MC~\cite{van2018graph},
    SpectralCF~\cite{zheng2018spectral},
    STAR-GCN~\cite{zhang2019star},
    NGCF~\cite{wang2019neural},\\
    Bi-HGNN~\cite{li2019hierarchical},
    LR-GCCF~\cite{chen2020revisiting},
    LightGCN~\cite{he2020lightgcn},
    AGCN~\cite{wu2020joint},\\
    HashGNN~\cite{tan2020learning},
    DGCN-BinCF~\cite{wang2019binarized}};
    \draw[] (n211.east) -- (n2111.west);

    \node [cate, below=1.6cm of n2111.west, anchor=west] (n2121) {\textbf{Adding Edges}:\\
    Multi-GCCF~\cite{sun2019multi}, DGCF~\cite{liu2020deoscillated}, DHCF~\cite{ji2020dual}};
    \node [cate, below=1.1cm of n2121.west, anchor=west] (n2122) {\textbf{Adding Nodes:}
    DGCF~\cite{wang2020disentangled}, HiGNN~\cite{li2020hierarchical2}};
    \draw[] (n212.east) -- (n2121.west);
    \draw[] (n212.east) -- (n2122.west);
    
    \node [cate, below=1.2cm of n2122.west, anchor=west] (n2131) {\textbf{Random-walk Sampling:}
    PinSage~\cite{ying2018graph}};
    \node [cate, below=1cm of n2131.west, anchor=west] (n2132) {\textbf{Random Sampling:}
    NIA-GCN~\cite{sun2020neighbor}, Multi-GCCF~\cite{sun2019multi}};
    \draw[] (n213.east) -- (n2131.west);
    \draw[] (n213.east) -- (n2132.west);
    
    \node [cate, below=2cm of n213.west, anchor=west] (n221) {\textbf{Equal Weights:}\\
    GC-MC~\cite{van2018graph}, STAR-GCN~\cite{zhang2019star}, Multi-GCCF~\cite{sun2019multi}, IC-MC~\cite{zhang2019inductive}, \\
    HashGNN~\cite{tan2020learning}, Bi-HGNN~\cite{li2019hierarchical},
    HiGNN~\cite{li2020hierarchical2}, GraphSAIL~\cite{xu2020graphsail}};
    \node [cate, below=1.5cm of n221.west, anchor=west] (n222) {\textbf{``Degree Normalization'':} \\
    LR-GCCF~\cite{chen2020revisiting}, AGCN~\cite{wu2020joint},
    LightGCN~\cite{he2020lightgcn}, DGCF~\cite{liu2020deoscillated}};
    \node [cate, below=1cm of n222.west, anchor=west] (n223) {\textbf{Attentive Weights:} 
    MCCF~\cite{wang2019multic},DisenHAN~\cite{wang2020disenhan}};
    \node [cate, below=1cm of n223.west, anchor=west] (n224) {\textbf{Central Node Augmentation:} NGCF~\cite{wang2019neural}};
    \node [cate, below=1cm of n224.west, anchor=west] (n225) {\textbf{Neighbor Interaction:} NIA-GCN~\cite{sun2020neighbor}};
    \foreach \x in {1,2,3,4,5}
    {\draw[] (n22.east) -- (n22\x.west);}

    \node [cate, below=1.5cm of n225.west, anchor=west] (n231) {$\mathbf{h}_u^{(l+1)}=\delta(\mathbf{W}^{(l)}\mathbf{n}_u^{(l)})$\\
    GC-MC~\cite{van2018graph}, STAR-GCN~\cite{zhang2019star},
    DGCF~\cite{wang2020disentangled}};
    \node [cate,  below=1.6cm of n231.west, anchor=west] (n232) {\textbf{Sum/Average of $\mathbf{h}_u^{(l)},\mathbf{n}_u^{(l)}$:}\\
    NGCF~\cite{wang2019neural}, IC-MC~\cite{zhang2019inductive}, HashGNN~\cite{tan2020learning},\\
    NIA-GCN~\cite{sun2020neighbor}, AGCN~\cite{wu2020joint}};
    \node [cate, below=1.8cm of n232.west, anchor=west] (n233) {$\textbf{h}_u^{(l+1)} = \sigma\left(\mathbf{W}^{(l)} \cdot (\mathbf{h}_{u}^{(l)}\oplus \mathbf{n}_{u}^{(l)})+\mathbf{b}^{(l)}\right)$\\
    PinSage~\cite{ying2018graph}, Bi-HGNN~\cite{li2019hierarchical}, GraphSAIL~\cite{xu2020graphsail},\\
    Multi-GCCF~\cite{sun2019multi}, HiGNN~\cite{li2020hierarchical2}};
    \node [cate, below=1.6cm of n233.west, anchor=west] (n234) {
    LightGCN~\cite{he2020lightgcn}: $\mathbf{h}_u^{(l+1)}=\mathbf{n}_u^{(l)}$\\
    LR-GCCF~\cite{chen2020revisiting}: $\mathbf{h}_u^{(l+1)}=\mathbf{W}^{(l)}\mathbf{n}_u^{(l)}$};
    \foreach \x in {1,2,3,4}
    {\draw[] (n23.east) -- (n23\x.west);}
    
    \node [cate, below=1.5cm of n234.west, anchor=west] (n241) {\textbf{Last Layer Representation} $\textbf{h}_u^*=\textbf{h}_u^{(L)}$};
    \node [cate, below=3.5cm of n241.west, anchor=west] (n242) {\textbf{Integration of All Layers}};
    \draw[] (n24.east) -- (n241.west);
    \draw[] (n24.east) -- (n242.west);
    
    \node [cate, right of=n241, node distance=7cm,] (n2411) {
    GC-MC~\cite{van2018graph}, PinSage~\cite{ying2018graph},\\
    STAR-GCN~\cite{zhang2019star}, Bi-HGNN~\cite{li2019hierarchical}, \\
    HashGNN~\cite{tan2020learning}, DisenHAN~\cite{wang2020disenhan}};
    \draw[] (n241.east) -- (n2411.west);
    
    \node [cate, below=1.8cm of n2411.west, anchor=west] (n2421) {$ \textbf{h}_u^*=\textbf{h}_u^{(0)}\oplus \textbf{h}_u^{(1)}\oplus\cdots\oplus \textbf{h}_u^{(L)}$\\
    SpectralCF~\cite{zheng2018spectral}, NGCF~\cite{wang2019neural}, IC-MC~\cite{zhang2019inductive}, \\ LR-GCCF~\cite{chen2020revisiting},
    MBGCN~\cite{jin2020multi}};
    \node [cate, below=1.3cm of n2421.west, anchor=west] (n2422) {
    $\textbf{h}_u^*=\frac{1}{L+1}\sum_{l=0}^{L}\mathbf{h}_u^{(l)}$
    DGCF~\cite{liu2020deoscillated}};
    \node [cate, below=1.0cm of n2422.west, anchor=west] (n2423) {
    $\textbf{h}_u^*=\sum_{l=0}^{L}\textbf{h}_u^{(l)}$
    DGCF~\cite{wang2020disentangled}};
    \node [cate, below=1.1cm of n2423.west, anchor=west] (n2424) {
    $\textbf{h}_u^*=\frac{1}{L+1}\sum_{l=0}^{L}\alpha^{(l)}\textbf{h}_u^{(l)}$\\
    LightGCN~\cite{he2020lightgcn}, GCM~\cite{wu2020graph}};
    \foreach \x in {1,2,3,4}
    {\draw[] (n242.east) -- (n242\x.west);}

    \end{tikzpicture}}
    \caption{\blue{Summary of representative works in user-item collaborative filtering.}}
    \label{fig:user-item-summ-fig}
\end{figure}
\subsection{Summary}
\blue{Corresponding to the discussion at the beginning of this section, we briefly summarize the existing works from four issues:}

    $\bullet$ \textbf{Graph Construction.} The most straightforward way is to directly use the original user-item bipartite graph.
    If some nodes have few neighbors in the original graph, it would be beneficial to enrich the graph structure by either adding edges or nodes.
    When dealing with large-scale graphs, it is necessary to sample the neighborhood for computational efficiency.
    Sampling is a trade-off between effectiveness and efficiency, and a more effective sampling strategy deserves further study.
    
    $\bullet$ \textbf{Neighbor Aggregation.} 
    When neighbors are more heterogeneous, aggregating neighbors with attentive weights would be preferable to equal weights and degree normalization; otherwise, the latter two are preferable for easier calculation. %
    Explicitly modeling the influence among neighbors or the affinity between the central node and neighbors might bring additional benefits, but needs to be verified on more datasets.
    
    $\bullet$ \textbf{Information Update.}
    Compared to discarding the original node, updating the node with its original representation and the aggregated neighbor representation would be preferable.
    Recent works show that simplifying the traditional GCN by removing the transformation and non-linearity operation can achieve better performance than the original ones.
    
    $\bullet$ \textbf{Final Node Representation.}
    To obtain overall user/item representation, utilizing the representations from all layers is preferable to directly using the last layer representation.
    In terms of the function of integrating the representations from all layers, weighted-pooling allows more flexibility, and concatenation preserves information from all layers.

Figure~\ref{fig:user-item-summ-fig} summarizes the typical strategies for each of the main issues, and lists the representative works accordingly.

\section{Sequential Recommendation}
Sequential recommendation predicts users’ next preferences based on their most recent activities, which seeks to model sequential patterns among successive items, and generate accurate recommendations for users~\cite{quadrana2018sequence}.
From the perspective of adjacency between items, sequences of items can be modeled as graph-structured data.   
Inspired by the advantage of GNN, it is becoming popular to utilize GNN to capture the transition pattern from users' sequential behaviors by transforming them into the sequence graph.

\tikzset{
  FARROW/.style={arrows={-{Latex[length=1.25mm, width=1.mm]}}, }, %
  U/.style = {circle, draw=melon!400, fill=melon, minimum width=1.4em, align=center, inner sep=0, outer sep=0},
  I/.style = {circle, draw=tea_green!400, fill=tea_green, minimum width=1.4em, align=center, inner sep=0, outer sep=0},
  cate/.style = {rectangle, draw, minimum width=8em, minimum height=2em, align=center, rounded corners=3}, %
  cate2/.style = {rectangle, minimum width=2em, minimum height=2em, align=center, rounded corners=3},%
  encoder/.style = {rectangle, fill=Madang!82, minimum width=10em, minimum height=3em, align=center, rounded corners=3},
  v_rep/.style={
       rectangle split,
       rectangle split part align=base,
       rectangle split horizontal=true,
       rectangle split draw splits=true,
       rectangle split parts=5,
       rectangle split part fill={red!30, blue!20, athens_gray!80, matisse, silver},
       draw=gray, %
       very thin,
       minimum height=1em,
       minimum width=2em,
       inner sep=2.5pt,
       text centered,
       text=gray,
       rounded corners=1
       },
}

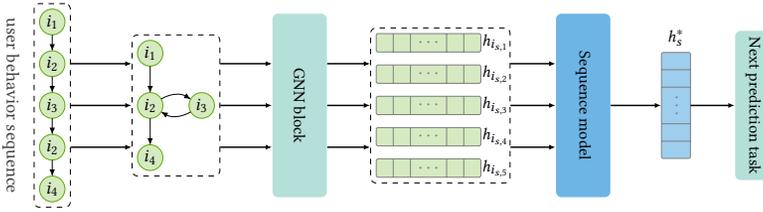
\begin{figure}
\centering
\resizebox{0.75\linewidth}{!}{
    \begin{tikzpicture}
        \node [I, node distance=4cm, yshift=-0.5cm] (is1) at (0, 0) {$i_{1}$};
     \node [I, below of=is1, node distance=0.8cm] (is2)  {$i_{2}$};
     \node [I, below of=is2, node distance=0.8cm] (is3) {$i_{3}$};
     \node [I, below of=is3, node distance=0.8cm] (is4) {$i_{2}$};
     \node [I, below of=is4, node distance=0.8cm] (is5) {$i_{4}$};
     \node [align=center, left of=is3, node distance=0.8cm, rotate=-90, gray!150] (ranking) {user behavior sequence};
     
    \node [I, right of=is3, node distance=1.9cm] (i2) {$i_{2}$};
    \node [I, above of=i2, node distance=1cm,] (i1) {$i_1$};
    \node [I, right of=i2, node distance=1cm,] (i3) {$i_3$};
    \node [I, below of=i2, node distance=1cm,] (i4) {$i_4$};
    \draw[FARROW] (i1) -> (i2);
    \draw[FARROW] (i2) -> (i4);
    \draw[FARROW] (i2) edge[bend left=30] (i3);
    \draw[FARROW] (i3) edge[bend right=-30] (i2);
    \draw[FARROW] (is1) -> (is2);
    \draw[FARROW] (is2) -> (is3);
    \draw[FARROW] (is3) -> (is4);
    \draw[FARROW] (is4) -> (is5);
    
    \node[encoder, fill=monte_carlo!62, right of=i3, node distance=1.9cm, align=center, rotate=-90] (sg) {\small GNN block};
    
    \node[v_rep, rectangle split part fill={tea_green}, right of=is3, node distance=7.3cm] (gi3) {
      \nodepart{three} $\,\cdots$
     };
     \node [right of=gi3, node distance=1.3cm] (h3) {\small $h_{i_{s,3}}$};
     \node[v_rep, rectangle split part fill={tea_green}, right of=is2, node distance=7.3cm, yshift=-0.2cm] (gi2) {
      \nodepart{three} $\,\cdots$
     };
    \node [right of=gi2, node distance=1.3cm] (h2) {\small $h_{i_{s,2}}$};
    \node[v_rep, rectangle split part fill={tea_green}, right of=is1, node distance=7.3cm, yshift=-0.4cm] (gi1) {
      \nodepart{three} $\,\cdots$
     };
     \node [right of=gi1, node distance=1.3cm] (h1) {\small $h_{i_{s,1}}$};
     \node[v_rep, rectangle split part fill={tea_green}, right of=is4, node distance=7.3cm, yshift=0.2cm] (gi4) {
      \nodepart{three} $\,\cdots$
     };
     \node [right of=gi4, node distance=1.3cm] (h4) {\small $h_{i_{s,4}}$};
     \node[v_rep, rectangle split part fill={tea_green}, right of=is5, node distance=7.3cm, yshift=0.4cm] (gi5) {
      \nodepart{three} $\,\cdots$
     };
     \node [right of=gi5, node distance=1.3cm] (h5) {\small $h_{i_{s,5}}$};
    \node[encoder, fill=celestial_blue!72, right of=sg, node distance=5.5cm, align=center, rotate=-90] (seq) {\small Sequence model};
    
    \node[v_rep, minimum height=1.6em, rectangle split part fill={celestial_blue!48}, right of=seq, node distance=1.8cm, rotate=-90] (seq_v) {
      \nodepart{three} $\,\cdots$
     };
     \node [above of=seq_v, node distance=1.3cm] (s) {\small $h_s^*$};
     \node [dashed, draw=black, fit={(is1) (is5)}, inner sep=3, rounded corners=3] (sf) {};
     
     \node[encoder, fill=monte_carlo!62, right of=seq_v, node distance=1.5cm, align=center, rotate=-90, minimum width=8em, minimum height=2em] (np) {\small Next prediction task};
     
     \node [dashed, draw=black, fit={(i1) (i3) (i4)}, inner sep=3, rounded corners=3] (gf) {};
     
     \node [dashed, draw=black, fit={(gi1) ([xshift=-0.2cm] h3.east) (gi5)}, inner sep=3, rounded corners=3] (ivf) {};

      \draw[FARROW, thick] (sf.east) -- (gf.west);
      \draw[FARROW, thick] ($(sf.east)+(0, -0.8)$) -- ($(gf.west)+(0, -0.8)$);
      \draw[FARROW, thick] ($(sf.east)+(0, 0.8)$) -- ($(gf.west)+(0, 0.8)$);
      \draw[FARROW, thick] (gf.east) -- (sg.south);
      \draw[FARROW, thick] ($(gf.east)+(0, -0.8)$) -- ($(sg.south)+(0, -0.8)$);
      \draw[FARROW, thick] ($(gf.east)+(0, 0.8)$) -- ($(sg.south)+(0, 0.8)$);
       \draw[FARROW, thick] (sg.north) -- (ivf.west);
      \draw[FARROW, thick] ($(sg.north)+(0, -0.8)$) -- ($(ivf.west)+(0, -0.8)$);
      \draw[FARROW, thick] ($(sg.north)+(0, 0.8)$) -- ($(ivf.west)+(0, 0.8)$);
      
      \draw[FARROW, thick] (ivf.east) -- (seq.south);
      \draw[FARROW, thick] ($(ivf.east)+(0, -0.8)$) -- ($(seq.south)+(0, -0.8)$);
      \draw[FARROW, thick] ($(ivf.east)+(0, 0.8)$) -- ($(seq.south)+(0, 0.8)$);
      \draw[FARROW, thick] (seq.north) -- (seq_v.south);
      \draw[FARROW, thick] (seq_v.north) -- (np.south);

    \end{tikzpicture}
    }
    \caption{The overall framework of GNN in sequential recommendation.}
    \label{fig:seq_model}
\end{figure}

Figure~\ref{fig:seq_model} illustrates the overall framework of GNN in sequential recommendation.
To fully utilize GNN in the sequential recommendation, there are three main issues to deal with:

    $\bullet$ \textbf{Graph Construction}. To apply GNN in the sequential recommendation, the sequence data should be transformed into a sequence graph.
    Is it sufficient to construct a subgraph for each sequence independently? 
    Would it be better to add edges among several consecutive items than only between the two consecutive items?
    
    $\bullet$ \textbf{Information Propagation}. To capture the transition patterns, which propagation mechanism is more appropriate?
    Is it necessary to distinguish the sequential order of the linked items?
    
    $\bullet$ \textbf{Sequential Preference}. To get the user's temporal preference, the item representations in a sequence should be integrated.
    Whether to simply apply attentive pooling or leverage RNN structure to enhance consecutive time patterns?

\subsection{Graph construction}
Unlike the user-item interactions which have essentially bipartite graph structure, the sequential behaviors are naturally expressed in the order of time, \emph{i.e.}, sequences, instead of sequence graphs.
Constructing graph based on the original bipartite graph is optional and mainly driven by the scalability or heterogeneity issue, whereas the construction of sequence graph based on users' sequential behaviors is a necessity for applying GNN in sequential recommendation.
Figure~\ref{fig:seq_graph_construction} shows the representative graph construction strategies for sequential behaviors.

\tikzset{
  FARROW/.style={arrows={-{Latex[length=1.25mm, width=1.mm]}}, }, %
  FARROW2/.style={arrows={-{Latex[length=1.25mm, width=1.mm]}}, draw=light_coral},
  U/.style = {circle, draw=light_coral!400, fill=light_coral, minimum width=1.4em, align=center, inner sep=0, outer sep=0},
  I/.style = {circle, draw=tea_green!400, fill=tea_green, minimum width=1.4em, align=center, inner sep=0, outer sep=0},
  I2/.style = {circle, draw=light_coral!400, fill=light_coral, minimum width=1.4em, align=center, inner sep=0, outer sep=0},
  cate/.style = {rectangle, minimum width=2em, minimum height=1.2em, align=center, rounded corners=3, font=\scriptsize}, %
}

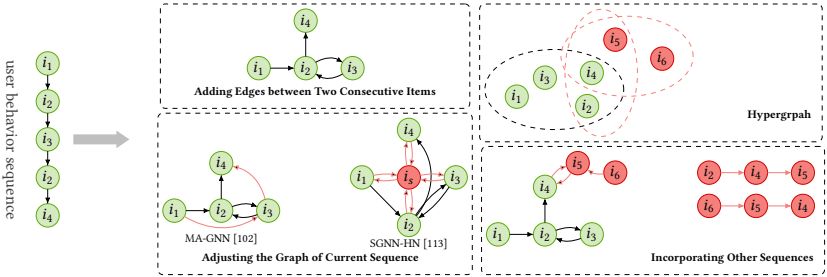
\begin{figure}
\resizebox{0.8\linewidth}{!}{
        \begin{tikzpicture}[]
    
     \node [I, node distance=4cm, yshift=-0.5cm] (is1) at (0, 0) {$i_{1}$};
     \node [I, below of=is1, node distance=0.8cm] (is2)  {$i_{2}$};
     \node [I, below of=is2, node distance=0.8cm] (is3) {$i_{3}$};
     \node [I, below of=is3, node distance=0.8cm] (is4) {$i_{2}$};
     \node [I, below of=is4, node distance=0.8cm] (is5) {$i_{4}$};
     \node [align=center, left of=is3, node distance=0.8cm, rotate=-90, gray!150] (ranking) {user behavior sequence};
     
    \node [I, right of=is3, node distance=4.5cm, yshift=1.5cm] (i1) {$i_{1}$};
    \node [I, right of=i1, node distance=1cm,] (i2) {$i_2$};
    \node [I, right of=i2, node distance=1cm,] (i3) {$i_3$};
    \node [I, above of=i2, node distance=1cm,] (i4) {$i_4$};
    \node [cate, below of = i2, node distance=0.5cm, xshift=0.2cm] (d1) {\quad\quad \textbf{Adding Edges between Two Consecutive Items}\quad\quad};
    \draw[FARROW] (i1) -> (i2);
    \draw[FARROW] (i2) -> (i4);
    \draw[FARROW] (i2) edge[bend left=30] (i3);
    \draw[FARROW] (i3) edge[bend right=-30] (i2);
    \draw[FARROW] (is1) -> (is2);
    \draw[FARROW] (is2) -> (is3);
    \draw[FARROW] (is3) -> (is4);
    \draw[FARROW] (is4) -> (is5);
    \node [dashed, draw=black, fit={(i4)(d1)}, inner sep=3, rounded corners=3] (m1) {};
    
    \draw[-{Triangle[width=10pt, length=8pt]}, line width=6pt, silver] ($(is3.east)+(0.3, 0)$) -- ++(1.2, 0);
    
    \node [I, right of=is3, node distance=2.7cm, yshift=-1.5cm] (i21) {$i_{1}$};
    \node [I, right of=i21, node distance=1cm,] (i22) {$i_2$};
    \node [I, right of=i22, node distance=1cm,] (i23) {$i_3$};
    \node [I, above of=i22, node distance=1cm,] (i24) {$i_4$};
    \node [cate, below of = i22, node distance=0.6cm] (f1) {MA-GNN~\cite{ma2019memory}};
    \node [cate, below of = i23, node distance=1cm, xshift=0.9cm] (d2) {\textbf{Adjusting the Graph of Current Sequence}};
    \draw[FARROW] (i21) -> (i22);
    \draw[FARROW] (i22) -> (i24);
    \draw[FARROW] (i22) edge[bend left=20] (i23);
    \draw[FARROW] (i23) edge[bend right=-20] (i22);
    \draw[FARROW2] (i21) edge[bend right=30] (i23);
    \draw[FARROW2] (i23) edge[bend right=30] (i24);
    
    \node [I, right of=i23, node distance=2.0cm, yshift=0.7cm] (i31) {$i_{1}$};
    \node [I2, right of=i31, node distance=1cm] (i35) {$i_s$};
    \node [I, below of=i35, node distance=1cm] (i32) {$i_2$};
    \node [I, right of=i35, node distance=1cm] (i33) {$i_3$};
    \node [I, above of=i35, node distance=1cm,] (i34) {$i_4$};
    \node [cate, below of = i32, node distance=0.4cm] (f2) {SGNN-HN~\cite{pan2020star}};
    \draw[FARROW] (i31) -> (i32);
    \draw[FARROW] (i32) edge[bend right=40] (i34);
    \draw[FARROW] (i32) edge[bend left=10] (i33);
    \draw[FARROW] (i33) edge[bend right=-10] (i32);
    \draw[FARROW2] (i31) edge[bend left=10] (i35);
    \draw[FARROW2] (i35) edge[bend left=10] (i31);
    \draw[FARROW2] (i32) edge[bend left=10] (i35);
    \draw[FARROW2] (i35) edge[bend left=10] (i32);
    \draw[FARROW2] (i33) edge[bend left=10] (i35);
    \draw[FARROW2] (i35) edge[bend left=10] (i33);
    \draw[FARROW2] (i34) edge[bend left=10] (i35);
    \draw[FARROW2] (i35) edge[bend left=10] (i34);
    
    \node [dashed, draw=black, fit={(i21)(i34)(i33)(d2)}, inner sep=3, rounded corners=3] (m2) {};
    
    \node [I, right of=i3, node distance=3.5cm, yshift=-0.6cm] (i41) {$i_{1}$};
    \node [I, right of=i41, node distance=1.5cm, yshift=-0.2cm] (i42) {$i_2$};
    \node [I, right of=i41, node distance=0.6cm, yshift=0.4cm] (i43) {$i_3$};
    \node [I, right of=i41, node distance=1.6cm, yshift=0.5cm] (i44) {$i_4$};
    \node [I2, right of=i44, node distance=0.5cm, yshift=0.7cm] (i45) {$i_5$};
    \node [I2, right of=i44, node distance=1.5cm, yshift=0.3cm] (i46) {$i_6$};
    \node[ellipse, dashed, draw=black, fit=(i41) (i42) (i43) (i44),inner sep=0pt] (e1) {};
    \node[ellipse, dashed, draw=light_coral, fit=(i42) (i44) (i45),inner sep=0pt] (e2) {};
    \node[ellipse, dashed, draw=light_coral, fit=(i44) (i45) (i46),inner sep=0pt] (e3) {};
    \node [cate, right of = i46, node distance=2.5cm, yshift=-1.2cm] (d3) {\textbf{Hypergrpah}};
    \node [dashed, draw=black, fit={(e1)(e2)(e3)(d3)}, inner sep=3, rounded corners=3] (m3) {};
    
    \node [I, right of=i3, node distance=3.1cm, yshift=-3.5cm] (i51) {$i_{1}$};
    \node [I, right of=i51, node distance=1cm] (i52) {$i_2$};
    \node [I, right of=i52, node distance=1cm] (i53) {$i_3$};
    \node [I, above of=i52, node distance=1cm] (i54) {$i_4$};
    \node [I2, right of=i54, node distance=0.7cm, yshift=0.5cm] (i55) {$i_5$};
    \node [I2, right of=i54, node distance=1.5cm, yshift=0.3cm] (i56) {$i_6$};
    \node [cate, right of = i53, node distance=3cm, yshift=-0.5cm] (d4) {\textbf{Incorporating Other Sequences}};
    \draw[FARROW] (i51) -> (i52);
    \draw[FARROW] (i52) -> (i54);
    \draw[FARROW] (i52) edge[bend left=20] (i53);
    \draw[FARROW] (i53) edge[bend right=-20] (i52);
    \draw[FARROW2] (i54) edge[bend right=-20] (i55);
    \draw[FARROW2] (i56) edge[bend right=-20] (i55);
    \draw[FARROW2] (i55) edge[bend right=-20] (i54);
    
    \node [I2, right of=i56, node distance=2cm] (i57) {$i_{2}$};
    \node [I2, right of=i57, node distance=1cm] (i58) {$i_{4}$};
    \node [I2, right of=i58, node distance=1cm] (i59) {$i_{5}$};
    \draw[FARROW2] (i57) -> (i58);
    \draw[FARROW2] (i58) -> (i59);
    \node [I2, below of=i57, node distance=0.7cm] (i60) {$i_{6}$};
    \node [I2, right of=i60, node distance=1cm] (i61) {$i_{5}$};
    \node [I2, right of=i61, node distance=1cm] (i62) {$i_{4}$};
    \draw[FARROW2] (i60) -> (i61);
    \draw[FARROW2] (i61) -> (i62);
    \node [dashed, draw=black, fit={(i51)(i55)(i62)(d4)}, inner sep=3, rounded corners=3] (m3) {};
    \end{tikzpicture}
}
\caption{\blue{Representative graph construction methods in the sequential recommendation, where the original nodes and edges are in green, and the additional ones are in red.}}
\label{fig:seq_graph_construction}
\end{figure}

\blue{Constructing the directed graph for each sequence by treating each item in the sequence as a node and adding edges between two consecutively clicked items is the most straightforward way~\cite{wu2019session,gupta2019niser,qiu2019rethinking,xu2019graph,qiu2020gag}.}
However, in most scenarios, the length of the user sequence is short, e.g., the average length on the preprocessed Yoochoose1/4\footnote{The dataset is available at http://2015.recsyschallenge.com/challege.
html. Note that this work preprocesses the dataset by filtering out the sequences of length 1 and items appearing less than 5 times.} dataset is 5.71~\cite{wu2019session}.
A sequence graph constructed from a single and short sequence consists of a small number of nodes and connections, and some nodes might even have only one neighbor, which contains too limited knowledge to reflect users' dynamic preferences and cannot take full advantage of GNN in graph learning.
To tackle this challenge, recent works propose several strategies to enrich the original sequence graph structure, which can be divided into two mainstreams.

\blue{One mainstream is to utilize additional sequences to enrich the item-item transitions.
The additional sequences can be other types of behavior sequences~\cite{wang2020beyond}, the historical sequences of the same user~\cite{wu2019personalizing}, or part/all of the sequences in the whole dataset~\cite{wang2020global,chen2021dual,zheng2020dgtn,zhou2021temporal}.
For instance, HetGNN~\cite{wang2020beyond} utilizes all behavior sequences, and constructs edges between two consecutive items in the same sequence with their behavior types as the edge types.
A-PGNN~\cite{wu2019personalizing} deals with the occasion when users are known, thus incorporating the user's historical sequences with the current sequence to enrich the item-item connections.
GCE-GNN~\cite{wang2020global} and DAT-MDI~\cite{chen2021dual} exploit the item transitions in all sessions to assist the transition patterns in the current sequence, which leverage the local context and global context.
Different from GCE-GNN~\cite{wang2020global} and DAT-MDI~\cite{chen2021dual} that treats all the transitions equally, TASRec~\cite{zhou2021temporal} attaches more importance to the recent transitions to augment the more recent transitions.
Instead of incorporating all the sessions, DGTN~\cite{zheng2020dgtn} only adds similar sessions to the current session, based on the assumption that similar sequences are more likely to reflect similar transition patterns.}
All these methods introduce more information into the original graph, and improve the performance compared to a single sequence graph.

Another mainstream approach is to adjust the graph structure of the current sequence. %
For example, assuming the current node has a direct influence on more than one consecutive item, MA-GNN~\cite{ma2019memory} extracts three subsequent items and adds edges between them.
Considering that only adding edges between consecutive items might neglect the relationships between distant items, SGNN-HN~\cite{pan2020star} introduces a virtual ``star'' node as the center of the sequence, which is linked with all the items in the current sequence.
The vector-wise representation of the ``star'' node reflects the overall characteristics of the whole sequence.
Hence, each item can gain some knowledge of the items without direct connections through the ``star'' node.
\citet{chen2020handling} point out that existing graph construction methods ignore the sequential information of neighbors, and bring about the ineffective long-term capturing problem.
Therefore, they propose LESSR, which constructs two graphs from one sequence: one distinguishes the order of neighbors, another allows the short-cut path from the item to all the items after it.

\blue{In addition to the above two mainstreams, other graph construction methods have emerged recently. 
Inspired by the advantage of hypergraph in modeling beyond-pairwise relations, hypergraph has been leveraged to capture the high-order relations among items and the cross-session information.
SHARE~\cite{wang2021session} constructs a hypergraph for each session, of which the hyperedges are defined by various sizes of sliding windows.
DHCN~\cite{xia2021self} takes each session as one hyperedge and integrates all the sessions in one hypergraph.
To explicitly incorporate cross-session relationships, DHCN~\cite{xia2021self} and COTREC~\cite{xia2021supervised} construct the session-to-session graph, which takes each session as a node and assigns the weights based on the shared items.}

\subsection{Information propagation}
Given a built sequence graph, it is essential to design an efficient propagation mechanism to capture transition patterns among items.
The GGNN framework is widely adopted to propagate information on the directed graph.
Specifically, it employs mean-pooling to aggregate the information of the previous items and the next items respectively, combines the two aggregated representations, and utilizes GRU\blue{~\cite{li2015gated}} component to integrate the information of neighbors and the central node.
The propagation functions are given as follows:
\begin{equation}
\begin{split}
    \mathbf{n}_{i_{s,t}}^{\mathrm{in}(l)}=\frac{1}{|\mathcal{N}_{i_{s,t}}^{\mathrm{in}}|}\Sigma_{j\in \mathcal{N}_{i_{s,t}}^{\mathrm{in}}}\mathbf{h}_j^{(l)}&,
    \quad 
    \mathbf{n}_{i_{s,t}}^{\mathrm{out}(l)}=\frac{1}{|\mathcal{N}_{i_{s,t}}^{\mathrm{out}}|}\Sigma_{j\in \mathcal{N}_{i_{s,t}}^{\mathrm{out}}}\mathbf{h}_j^{(l)},\\
    \mathbf{n}_{i_{s,t}}^{(l)}=\mathbf{n}_{i_{s,t}}^{\mathrm{in}(l)}\oplus \mathbf{n}_{i_{s,t}}^{\mathrm{out}(l)}&,
    \quad \mathbf{h}_{i_{s,t}}^{(l+1)}=\textbf{GRU}(\mathbf{h}_{i_{s,t}}^{(l)},\mathbf{n}_{i_{s,t}}^{(l)}),
\end{split}
\end{equation}
where $\mathcal{N}_{i_{s,t}}^{\mathrm{in}}$, $\mathcal{N}_{i_{s,t}}^{\mathrm{out}}$ denotes the neighborhood set of previous items and next items, $\textbf{GRU}(\cdot)$ represents the GRU component.
Different from the pooling operation, the gate mechanism in GRU decides what information to be preserved and discarded.
Unlike GGNN which treats the neighbors equally, attention mechanism is also utilized to differentiate the importance of neighbors~\cite{qiu2019rethinking,wang2020global,chang2021sequential}.
All the above methods adopt the permutation-invariant aggregation function during the message passing, ignoring the order of items within the neighborhood, which may lead to the loss of information~\cite{chen2020handling}.
To address this issue, LESSR~\cite{chen2020handling} preserves the order of items in the graph construction, and leverages the GRU component\blue{~\cite{li2015gated}} to aggregate the neighbors sequentially, as the following equation:
\begin{equation}
    \mathbf{n}_{i_{s,t},k}^{(l)}=\textbf{GRU}^{(l)}(\mathbf{n}_{i_{s,t},k-1}^{(l)}, \mathbf{h}_{i_{s,t},k}^{(l)}),
\end{equation}
where $\mathbf{h}_{i_{s,t},k}^{(l)}$ represents the $k^{th}$ item in the neighborhood of $i_{s,t}$ ordered by time, and $\mathbf{n}_{i_{s,t},k}^{(l)}$ denotes the neighborhood representation after aggregating $k$ items.

\blue{For the sequence graph with hypergraph structure, DHCN~\cite{xia2021self} adopts the typical hypergraph neural network HGNN~\cite{feng2019hypergraph}, which treats the nodes equally during propagation.
To differentiate the importance of items within the same hyperedge, SHARE~\cite{wang2021session} designs two attention mechanisms to propagate the information of item nodes. One is the hyperedges, and the other is the information of the hyperedges to the connected item nodes.
For user-aware sequential recommendation, A-PGNN~\cite{wu2019personalizing} and GAGA~\cite{qiu2020gag} implicitly incorporate the user information, and augment the representations of items in the neighborhood with user representation.}

\subsection{Sequential preference}

Due to the limited iteration of propagation, GNN cannot effectively capture long-range dependency among items~\blue{\cite{chen2020handling}}.
Therefore, the representation of the last item (or any item) in the sequence is not sufficient enough to reflect the user's sequential preference.
Besides, most of the graph construction methods of transforming sequences into graphs lose part of the sequential information~\cite{chen2020handling}.
In order to obtain the effective sequence representation, existing works propose several strategies to integrate the item representations in the sequence.

Considering that the items in a sequence have different levels of priority, the attention mechanism is widely adopted for integration.
Some works~\cite{wu2019session,qiu2020gag,pan2020star,zheng2020dgtn} calculate the attentive weights between the last item and all the items in the sequence and aggregate the item representations as the global preference, and incorporate it with local preference (\emph{i.e.}, the last item representation) as the overall preference.
In this way, the overall preference relies heavily on the relevance of the last item to the user preference.
Inspired by the superiority of multi-layer self-attention strategy in sequence modeling, GC-SAN~\cite{xu2019graph} stacks multiple self-attention layers on the top of the item representations generated by GNN to capture long-range dependencies.

In addition to leveraging attention mechanism for sequence integration, sequential signals are explicitly incorporated into the integration process.
For instance, NISER~\cite{gupta2019niser} and GCE-GNN~\cite{wang2020global} add the positional embeddings, which reflect the relative order of the items, to effectively obtain position-aware item representations.
To balance the consecutive time and flexible transition pattern, FGNN~\cite{qiu2019rethinking} employs the GRU with attention mechanism to iteratively update the user preference with item representations in the sequence.

\blue{All the above works integrate the item representations within the user's behavior sequence to generate the representation of sequential preference.
Apart from these methods, DHCN~\cite{xia2021self} and COTREC~\cite{xia2021supervised} enrich the sequence graph by the session-to-session graph in the graph construction step. Therefore, they combine the sequential representation learned from the session-to-session graph and the one aggregated from items at this step.}

\tikzset{
  FARROW/.style={arrows={-{Latex[length=1.25mm, width=1.mm]}}, }, %
  U/.style = {circle, draw=melon!400, fill=melon, minimum width=1.4em, align=center, inner sep=0, outer sep=0},
  I/.style = {circle, draw=tea_green!400, fill=tea_green, minimum width=1.4em, align=center, inner sep=0, outer sep=0},
  cate/.style = {rectangle, draw, minimum width=8em, minimum height=2em, align=center, rounded corners=3}, %
  cate2/.style = {rectangle, minimum width=2em, minimum height=2em, align=center, rounded corners=3},%
  encoder/.style = {rectangle, fill=Madang!82, minimum width=10em, minimum height=3em, align=center, rounded corners=3},
  v_rep/.style={
       rectangle split,
       rectangle split part align=base,
       rectangle split horizontal=true,
       rectangle split draw splits=true,
       rectangle split parts=5,
       rectangle split part fill={red!30, blue!20, athens_gray!80, matisse, silver},
       draw=gray, %
       very thin,
       minimum height=1em,
       minimum width=2em,
       inner sep=2.5pt,
       text centered,
       text=gray,
       rounded corners=1
       },
}

\begin{figure}
    \centering
    \resizebox{0.9\linewidth}{!}{
    \begin{tikzpicture}
    \node [cate, distance=4cm, yshift=-0.5cm] (n1) at (0, 0) {\textbf{Sequential Recommendation}};
    \node [cate, right of=n1, node distance=6cm, yshift=5cm] (n21) {\textbf{Graph Construction}};
    \node [cate, below=5cm of n21.west, anchor=west] (n22) {\textbf{Information Propagation}};
    \node [cate, below=4cm of n22.west, anchor=west] (n23) {\textbf{Sequential Preference}};
    \foreach \x in {1,2,3}
    {\draw[] (n1.east) -- (n2\x.west);}
    
    \node [cate, right of=n21, node distance=9cm, yshift=2.5cm, align=left] (n211) {\textbf{Adding Edges between Two Consecutive Items:}\\
    SR-GNN~\cite{wu2019session}, GC-SAN~\cite{xu2019graph},
    NISER~\cite{gupta2019niser}, FGNN~\cite{qiu2019rethinking}, 
    GAG~\cite{qiu2020gag}};
    \node [cate, below=1.4cm of n211.west, anchor=west, align=left] (n212) {\textbf{Incorporating Other Sequences:}\\
    HetGNN~\cite{wang2020beyond}, A-PGNN~\cite{wu2019personalizing}, DGTN~\cite{zheng2020dgtn},
    GCE-GNN~\cite{wang2020global}, \\
    DAT-MDI~\cite{chen2021dual}, TASRec~\cite{zhou2021temporal}, COTREC~\cite{xia2021supervised}};
    \node [cate, below=1.45cm of n212.west, anchor=west, align=left] (n213) {\textbf{Adjusting the Graph of Current Sequence:}\\
    MA-GNN~\cite{ma2019memory}, SGNN-HN~\cite{pan2020star}, LESSR~\cite{chen2020handling}, SURGE~\cite{chang2021sequential}};
    \node [cate, below=1.0cm of n213.west, anchor=west, align=left] (n214) {\textbf{Hypergraph:} SHARE~\cite{wang2021session}, DHCN~\cite{xia2021self}};
    \foreach \x in {1,2,3,4}
    {\draw[] (n21.east) -- (n21\x.west);}
    
     \node [cate, below=1.5cm of n214.west, anchor=west, align=left] (n221) {
     \textbf{Variants of GGNN:}\\
     SR-GNN~\cite{wu2019session}, GC-SAN~\cite{xu2019graph}, 
     NISER~\cite{gupta2019niser}, SGNN-HN~\cite{pan2020star}, 
     DAT-MDI~\cite{chen2021dual}};
     \node [cate, below=1.2cm of n221.west, anchor=west, align=left] (n222) {\textbf{Attention Mechanism:}\\
     FGNN~\cite{qiu2019rethinking}, GCE-GNN~\cite{wang2020global},
     SURGE~\cite{chang2021sequential}};
     \node [cate, below=1.0cm of n222.west, anchor=west, align=left] (n223) {\textbf{GCN Framework:}
     TASRec~\cite{zhou2021temporal}, COTREC~\cite{xia2021supervised}};
     \node [cate, below=1.0cm of n223.west, anchor=west, align=left] (n224) {\textbf{GRU for Aggregation:} LESSR~\cite{chen2020handling}};
     \foreach \x in {1,2,3,4}
     {\draw[] (n22.east) -- (n22\x.west);}
     
     \node [cate, below=1.2cm of n224.west, anchor=west, align=left] (n231) {\textbf{Attention Mechanism:}\\
     SR-GNN~\cite{wu2019session}, GAG~\cite{qiu2020gag}, SGNN-HN~\cite{pan2020star},
     DGTN~\cite{zheng2020dgtn}, GC-SAN~\cite{xu2019graph}};
     \node [cate, below=1.2cm of n231.west, anchor=west, align=left] (n232) {
     \textbf{Adding Positional Embedding:}\\
     NISER~\cite{gupta2019niser}, GCE-GNN~\cite{wang2020global}};
     \node [cate, below=1.0cm of n232.west, anchor=west, align=left] (n233) {
     \textbf{GRU with Attention:} FGNN~\cite{qiu2019rethinking}};
     \foreach \x in {1,2,3}
    {\draw[] (n23.east) -- (n23\x.west);}
      
    \end{tikzpicture}}
    \caption{\blue{Summary of representative works in sequential recommendation.}}
    \label{fig:seq-summ-fig}
\end{figure}
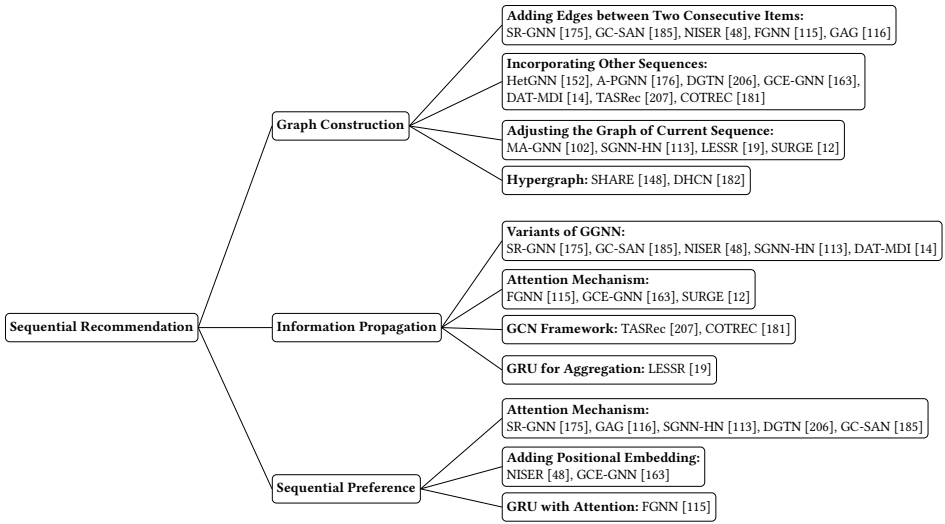
\subsection{Summary}
\blue{This part briefly summarizes the reviewed works in terms of the three main issues.}

    $\bullet$ \textbf{Graph Construction.} The most straightforward construction is to add edges between the two consecutive items. When the sequence length is short, utilizing additional sequences can enrich the sequence graph, and it would be preferable if the additional sequence are more similar to the original sequence.
    Another line is to adjust the graph structure of the behavior sequence. There is no accepted statement on which method is better.
    Moreover, incorporating the session-to-session graph into the sequence graph is also used to gain further improvements.
    
    $\bullet$ \textbf{Information Propagation.} Most of the propagation methods are variants of the propagation methods in traditional GNN frameworks, and there is no consensus on which method is better. %
    Some complex propagation methods, such as LESSR~\cite{chen2020handling}, achieve performance gain at the cost of more computation.
    Whether to adopt complex propagation methods in practice depends on the trade-off between computation costs and performance gains.
    
    $\bullet$ \textbf{Sequential Preference.} To obtain the sequential preference, an attention mechanism is widely adopted to integrate the representations of items in the sequence.
    Beyond that, adding positional embeddings can enhance the relative order of the items and can bring a few improvements. 
    Whether leveraging RNN structure can boost performance for all the sequential recommendation tasks requires further investigation.
    
Figure~\ref{fig:seq-summ-fig} summarizes the typical strategies for each of the main issues and lists the representative works accordingly.

\section{Social Recommendation}
With the emergence of online social networks, social recommender systems have been proposed to utilize each user’s local neighbors’ preferences to enhance user modeling~\cite{ma2008sorec,jamali2010matrix,ma2009learning,guo2015trustsvd,wu2019neural}. 
All these works assume users with social relationships should have similar representations based on the social influence theory that connected people would influence each other.
Some of them directly use such relationship as regularizer to constraint the final user representations~\cite{ma2008sorec,jamali2010matrix,ma2011recommender,tang2013exploiting},
while others leverage such relationship as input to enhance the original user embeddings~\cite{ma2009learning,guo2015trustsvd}.

From the perspective of graph learning, the early works mentioned above can be seen as modeling the first-order neighbors of each user.
However, in practice, a user might be influenced by her/his friends' friends.
Overlooking the high-order influence diffusion in previous works might lead to the suboptimal recommendation performance~\cite{wu2019neural}.
Thanks to the ability of simulating how users are influenced by the recursive social diffusion process, GNN has become a popular choice to model the social information in recommendation.

To incorporate relationships among users into interaction behaviors by leveraging GNN, there are two main issues to deal with:

    $\bullet$ \textbf{Influence of Friends}. 
    Do friends have equal influence? If not, how to distinguish the influence of different friends? 
    
    $\bullet$ \textbf{Preference Integration}. Users are involved in two types of relationships, \emph{i.e.}, social relationships with their friends and interactions with items.
    How to integrate the user representations from the social influence perspective and interaction behavior? 

\tikzset{
  FARROW/.style={arrows={-{Latex[length=1.25mm, width=1.mm]}}, }, %
  U/.style = {circle, draw=melon!400, fill=melon, minimum width=1.4em, align=center, inner sep=0, outer sep=0},
  I/.style = {circle, draw=tea_green!400, fill=tea_green, minimum width=1.4em, align=center, inner sep=0, outer sep=0},
  cate/.style = {rectangle, draw, minimum width=3em, minimum height=1.5em, align=center, rounded corners=3},
  encoder/.style = {rectangle, fill=Madang!82, minimum width=10em, minimum height=3em, align=center, rounded corners=3},
  cate2/.style = {rectangle, minimum width=1em, minimum height=2em, align=center, rounded corners=3},%
  v_rep/.style={
       rectangle split,
       rectangle split part align=base,
       rectangle split horizontal=true,
       rectangle split draw splits=true,
       rectangle split parts=5,
       rectangle split part fill={red!30, blue!20, athens_gray!80, matisse, silver},
       draw=gray, %
       very thin,
       minimum height=1.6em,
       minimum width=2.4em,
       inner sep=2.5pt,
       text centered,
       text=gray,
       rounded corners=1
       },
}

\begin{figure}
    \centering
    \resizebox{0.85\linewidth}{!}{
    \begin{subfigure}[b]{0.45\textwidth}
    \centering
    \resizebox{0.95\linewidth}{!}{
        \begin{tikzpicture}[]

    \node [U, align=center] (u2) at (0,0) {$u_2$};
    \node [U, left of=u2, node distance=0.75cm, align=center] (u1) {$u_1$};
    \node [U, right of=u2, node distance=0.75cm, align=center] (u3) {$u_3$};
    \node [U, right of=u3, node distance=0.75cm, align=center] (u4) {$u_4$};
    
     \foreach \x in {1,2,3,4}
    {
        \node [I, below of=u\x, node distance=1.2cm, xshift=-0.37cm] (i\x) {$i_{\x}$};
    } 
    \node [I, right of=i4, node distance=0.75cm,] (i5) {$i_{5}$};
    \draw[] (u1) -- (i2);
    \draw[] (u1) -- (i3) -- (u2) -- (i5) -- (u4);
    \draw[] (u3) -- (i1) -- (u4) -- (i3);
    \draw (u4) -- (i4) -- (u3);
    
    \node [U, right of=u4, node distance=3.5cm, yshift=-0.5cm] (uu1) {$u_1$};
    \node [U, above of=uu1, node distance=0.75cm, xshift=0.75cm] (uu2) {$u_2$};
    \node [U, right of=uu1, node distance=1.50cm, align=center] (uu3) {$u_3$};
    \node [U, below of=uu1, node distance=0.75cm, xshift=0.75cm] (uu4) {$u_4$};
    
    \node [dashed, draw=black, fit={([yshift=-0.5mm, xshift=-0.5mm] uu1.south west) ([yshift=0.5mm, xshift=0.5mm] uu3.north east) ([yshift=-0.5mm, xshift=0.5mm] uu4.south east) (uu2)}, inner sep=3, rounded corners=3] (o_s) {};
        
        \draw[] (uu1) -- (uu2);
        \draw[] (uu2) -- (uu3);
        \draw[] (uu3) -- (uu4);
        \draw[] (uu3) -- (uu1);

    \node[encoder, fill=flamingo!62, above of=uu2, node distance=1.8cm, align=center, minimum width=8em] (sg) {\small Social GNN block};

    \node [encoder, left of=sg, node distance=5.35cm, fill=monte_carlo!62] (uig) {\small U-I GNN block};
    
    \node [dashed, draw=black, fit={([yshift=-0.5mm, xshift=-0.5mm] i1.south west) ([yshift=0.5mm, xshift=0.5mm] u4.north east) ([yshift=-0.5mm, xshift=0.5mm] i5.south east)}, inner sep=3, rounded corners=3] (o_m) {};
     \node[v_rep, rectangle split part fill={melon}, above of=uig, node distance=1.5cm, xshift=2.2cm] (hui) {
      \nodepart{three} $\,\cdots$
     };
      \node[v_rep, rectangle split part fill={tea_green}, above of=uig, node distance=2.8cm] (hi) {
      \nodepart{three} $\,\cdots$
     };
     \node [right of=hi, node distance=1.3cm] (hi_a) {\small $h_i^*$};
     \node[v_rep, rectangle split part fill={flamingo!62}, above of=sg, node distance=1.5cm] (hus) {
      \nodepart{three} $\,\cdots$
     };
     \node [right of=hus, node distance=1.3cm] (hus_a) {\small $h_u^S$};
     
     \node[v_rep, rectangle split part fill={melon!80!flamingo}, above of=hui, node distance=1.3cm, xshift=2.2cm] (hs) {
      \nodepart{three} $\,\cdots$
     };
     \node [left of=hs, node distance=1.3cm] (hs_a) {\small $h_u^*$};
     
     \node [above of=hs, node distance=1.3cm, xshift=-2cm] (y) {$y_{ui}$};
     
     \node [right of=hui, node distance=1.3cm] (hui_a) {\small $h_u^I$};
     
     \draw[FARROW] ($(o_m.north)+(0, 0)$) -- ++(0, 1.2)  ; 
     \draw[FARROW] ($(o_m.north)+(-1, 0)$) -- ++(0, 1.2)  ;   
     \draw[FARROW] ($(o_m.north)+(1, 0)$) -- ++(0, 1.18)  ;
     \draw[FARROW] ($(o_s.north)+(0, 0)$) -- ++(0, 0.9)  ; 
     \draw[FARROW] ($(o_s.north)+(-0.8, 0)$) -- ++(0, 0.9)  ;   
     \draw[FARROW] ($(o_s.north)+(0.8, 0)$) -- ++(0, 0.9)  ; 
     \draw[FARROW] (uig.north) -> (hui.south) ; 
     \draw[FARROW] (sg.north) -> (hus.south) ; 
     \draw[FARROW] (uig.north) -> (hi.south) ; 
     \draw[FARROW] (hui.north) -> (hs.south) ; 
     \draw[FARROW] (hus.north) -> (hs.south) ; 
    
    \draw[FARROW, rounded corners,] (hi.north) |- (y.west) {} ;
    \draw[FARROW, rounded corners,] (hs.north) |- (y.east) {} ;
            
        \end{tikzpicture}
        }
        \caption{The framework of GNN on the bipartite graph and social network graph separately. }
        \label{fig:socialA_model}
    \end{subfigure}
\hspace{1cm}
    \begin{subfigure}[b]{0.45\textwidth}
    \centering
        \resizebox{0.95\linewidth}{!}{
        
        \begin{tikzpicture}[]
        
        \node [U, align=center] (u2) at (0,0) {$u_2$};
    \node [U, left of=u2, node distance=0.75cm, align=center] (u1) {$u_1$};
    \node [U, right of=u2, node distance=0.75cm, align=center] (u3) {$u_3$};
    \node [U, right of=u3, node distance=0.75cm, align=center] (u4) {$u_4$};
    
     \foreach \x in {1,2,3,4}
    {
        \node [I, below of=u\x, node distance=1.2cm, xshift=-0.37cm] (i\x) {$i_{\x}$};
    } 
    \node [I, right of=i4, node distance=0.75cm,] (i5) {$i_{5}$};
    \draw[] (u1) -- (i2);
    \draw[] (u1) -- (i3) -- (u2) -- (i5) -- (u4);
    \draw[] (u3) -- (i1) -- (u4) -- (i3);
    \draw (u4) -- (i4) -- (u3);
    \node [dashed, draw=black, fit={([yshift=-0.5mm, xshift=-0.5mm] i1.south west) ([yshift=0.5mm, xshift=0.5mm] u4.north east) ([yshift=-0.5mm, xshift=0.5mm] i5.south east)}, inner sep=3, rounded corners=3] (o_m) {};
    
     \node [U, right of=u4, node distance=3.5cm, yshift=-0.5cm] (uu1) {$u_1$};
    \node [U, above of=uu1, node distance=0.75cm, xshift=0.75cm] (uu2) {$u_2$};
    \node [U, right of=uu1, node distance=1.50cm, align=center] (uu3) {$u_3$};
    \node [U, below of=uu1, node distance=0.75cm, xshift=0.75cm] (uu4) {$u_4$};
    
    \node [dashed, draw=black, fit={([yshift=-0.5mm, xshift=-0.5mm] uu1.south west) ([yshift=0.5mm, xshift=0.5mm] uu3.north east) ([yshift=-0.5mm, xshift=0.5mm] uu4.south east) (uu2)}, inner sep=3, rounded corners=3] (o_s) {};
        
        \draw[] (uu1) -- (uu2);
        \draw[] (uu2) -- (uu3);
        \draw[] (uu3) -- (uu4);
        \draw[] (uu3) -- (uu1);

     \node [U, above of=uu2, node distance=1.6cm, xshift=-4cm] (cu1) {$u_1$};
    \node [U, above of=cu1, node distance=0.75cm, xshift=0.75cm] (cu2) {$u_2$};
    \node [U, right of=cu1, node distance=1.50cm, align=center] (cu4) {$u_4$};
    \node [U, below of=cu1, node distance=0.75cm, xshift=0.75cm] (cu3) {$u_3$};
    
    \node [I, left of=cu1, node distance=1cm,] (ci2) {$i_{2}$};
    \node [I, right of=cu3, node distance=1cm,] (ci1) {$i_{1}$};
     \node [I, above of=cu2, node distance=0.5cm, xshift=-1cm] (ci3) {$i_{3}$};
     \node [I, right of=cu4, node distance=1cm,] (ci4) {$i_{4}$};
     \node [I, right of=cu4, node distance=0.4cm, yshift=1cm] (ci5) {$i_{5}$};
     
       \draw[draw=melon!400,thick] (cu1) -- (cu2);
        \draw[draw=melon!400, thick] (cu2) -- (cu3);
        \draw[draw=melon!400, thick] (cu3) -- (cu4);
        \draw[draw=melon!400, thick] (cu3) -- (cu1);
        \draw[] (cu3) -- (ci1) -- (cu4);
        \draw[] (cu1) -- (ci2);
        \draw[] (cu1) -- (ci3) -- (cu2) -- (ci3);
        \draw[] (ci3) to [out=0,in=90] (cu4);
        \draw[] (cu3) -- (ci4) -- (cu4);
        \draw[] (cu2) -- (ci5) -- (cu4);
        
     \node [dashed, draw=black, fit={(ci2) (ci3)(ci4) (ci1) (cu3)}, inner sep=3, rounded corners=3] (o_c) {};
     
     \draw[FARROW, rounded corners] ($(o_m.north)+(-0.8, 0)$) |- (o_c.west)  ;
     \draw[FARROW, rounded corners] ($(o_s.north)$) |- (o_c.east)  ;
     
     \node[encoder, fill=flamingo!62, above of=o_c, node distance=2.35cm, align=center] (cg) {\small GNN block};
     
     \node[v_rep, rectangle split part fill={melon}, above of=cg, node distance=1.2cm, xshift=1.5cm] (hu) {
      \nodepart{three} $\,\cdots$
     };
     \node [right of=hu, node distance=1.3cm] (hu_1) {\small $h_u^*$};
     \node[v_rep, rectangle split part fill={tea_green}, above of=cg, node distance=1.2cm, xshift=-1.5cm] (hi) {
      \nodepart{three} $\,\cdots$
     };
     \node [left of=hi, node distance=1.3cm] (hi_1) {\small $h_i^*$};
     \draw[FARROW] ($(hi.south)+(0, -0.4)$) -- (hi.south)  ;
     \draw[FARROW] ($(hu.south)+(0, -0.4)$) -- (hu.south)  ;
     \draw[FARROW] ($(o_c.north)+(-1, 0)$) -- ($(cg.south)+(-1, 0)$) ;
     \draw[FARROW] ($(o_c.north)+(1, 0)$) --($(cg.south)+(1, 0)$) ;
     \draw[FARROW] ($(o_c.north)$) -- (cg.south)  ;
     
     \node [above of=cg, node distance=2cm] (y) {$y_{ui}$};
      \draw[FARROW, rounded corners,] (hi.north) |- (y.west) {} ;
    \draw[FARROW, rounded corners,] (hu.north) |- (y.east) {} ;
     
        \end{tikzpicture}
        }
        \caption{The framework of GNN on the unified graph of user-item interactions and social network. }
        \label{fig:socialB_model}
    \end{subfigure}}

    \caption{Two strategies for social enhanced general recommendation.}
    \label{fig:social_model}
    
\end{figure}
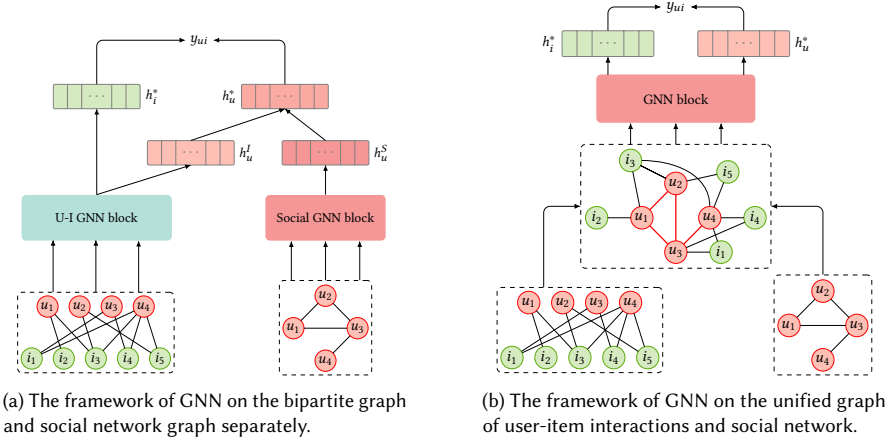

\subsection{Influence of friends}

\blue{Generally, a social graph only contains information about whether the users are friends, but the strengths of social ties are usually unknown.}
To propagate the information of friends, it is essential to decide the influence of friends.
DiffNet~\cite{wu2019neural} treats the influence of friends equally by leveraging mean-pooling operation.
However, the assumption of equal influence is not in accordance with the actual situation, and the influence of a user is unsuitable to be simply determined by the number of her/his friends.
Indeed, users are more likely to be influenced by friends with strong social ties or similar preferences.
Therefore, attention mechanism is widely leveraged to differentiate the influence of neighbors~\cite{fan2019graph,wu2019dual,wu2020diffnet++,GAT-NSR,song2019session,TGRec}.
\blue{For example, \citet{song2019session} propose DGRec, which dynamically infers the influence of neighbors based on their current interests.
It first models dynamic users' behaviors with a recurrent neural network, and then acquires the social influence with a graph-attention neural network.}
Compared to the mean-pooling operation, the attention mechanism boosts the overall performance, which further verifies the assumption that different friends have different influence power.

Moreover, a recent work, named ESRF~\cite{yu2020enhance}, argues that social relations are not always reliable.
The unreliability of social information lies in two aspects:
on the one hand, the users with explicit social connections might have no influence power;
on the other hand, the obtained social relationships might be incomplete.
Considering that indiscriminately incorporating unreliable social relationships into recommendation may lead to poor performance,
ESRP leverages the autoencoder mechanism to modify the observed social relationships by filtering irrelevant relationships and investigating the new neighbors.
\blue{Similarly, DiffNetLG~\cite{DiffNetLG} involves implicit local influence to predict the unobserved social relationship and then utilizes both explicit and implicit social relations to make recommendations.}

\subsection{Preference Integration}
Users in social recommendatioin are involved in two types of relationships, one is the user-item interactions and the other is the social graph.
To enhance the user preference representation by leveraging social information, there are two strategies for combining the information from these two networks, %

    $\bullet$ to learn the user representation from these two networks respectively~\cite{wu2019neural,fan2019graph,wu2019dual} and then integrate them into the final preference vector, as illustrated in Figure~\ref{fig:socialA_model};
    
    $\bullet$ to combine the two networks into one unified network~\cite{wu2020diffnet++} and apply GNN to propagate information, as illustrated in Figure~\ref{fig:socialB_model}.

The advantage of the first strategy lies in two aspects:
on the one hand, we can differentiate the depth of diffusion process of two networks since they are treated separately;
on the other hand, any advanced method for user-item bipartite graph can be directly applied, and for social network, a homogeneous graph, GNN techniques are extremely suitable for simulating the influence process since they are originally proposed for homogeneous graphs.
\blue{As for the integration of the user representations learned from two relationships, there are two main mechanisms, \emph{i.e.}, linearity combination and non-linearity combination.
Among the linearity combination, DiffNet~\cite{wu2019neural} treats the user representations from two spaces equally, and combines them with a sum-pooling operation.
Instead of an equal-weight combination, DANSER~\cite{wu2019dual} dynamically allocates weights according to the user-item paired features.
Among the non-linearity combination, multi-layer perceptrons over the concatenated vector are widely adopted to enhance the feature interactions~\cite{fan2019graph,GNN-SoR,SR-HGNN}.}

The advantage of integrating the two graphs into one unified network is that both the higher-order social influence diffusion in the social network
and interest diffusion in the user-item bipartite graph can be simulated in a unified model, and these two kinds of information simultaneously reflect users' preferences.
DiffNet++~\cite{wu2020diffnet++} designs a two-level attention network to update user nodes at each layer.
Specifically, it firstly aggregates the information of neighbors in the bipartite graph (\emph{i.e.}, interacted items) and social network (\emph{i.e.}, friends) by utilizing the GAT mechanism respectively.
Considering that different users may have different preferences in balancing these two relationships, it further leverages another attention network to fuse the two hidden states of neighbors.
\blue{Similarly, SEFrame~\cite{SEFrame} utilizes a heterogeneous graph network to fuse the knowledge from social relationships, user-item interactions and item transitions from the heterogeneous graph, and employs a two-level attention network for propagation.
}
Up till now, there is no evidence to show which strategy always achieves better performance.

\tikzset{
  FARROW/.style={arrows={-{Latex[length=1.25mm, width=1.mm]}}, }, %
  U/.style = {circle, draw=melon!400, fill=melon, minimum width=1.4em, align=center, inner sep=0, outer sep=0},
  I/.style = {circle, draw=tea_green!400, fill=tea_green, minimum width=1.4em, align=center, inner sep=0, outer sep=0},
  cate/.style = {rectangle, draw, minimum width=8em, minimum height=2em, align=center, rounded corners=3}, %
  cate2/.style = {rectangle, minimum width=2em, minimum height=2em, align=center, rounded corners=3},%
  encoder/.style = {rectangle, fill=Madang!82, minimum width=10em, minimum height=3em, align=center, rounded corners=3},
  v_rep/.style={
       rectangle split,
       rectangle split part align=base,
       rectangle split horizontal=true,
       rectangle split draw splits=true,
       rectangle split parts=5,
       rectangle split part fill={red!30, blue!20, athens_gray!80, matisse, silver},
       draw=gray, %
       very thin,
       minimum height=1em,
       minimum width=2em,
       inner sep=2.5pt,
       text centered,
       text=gray,
       rounded corners=1
       },
}

\begin{figure}
    \centering
    \resizebox{0.9\linewidth}{!}{
    \begin{tikzpicture}
    \node [cate, distance=4cm, yshift=-0.5cm] (n1) at (0, 0) {\textbf{Social Recommendation}};
    \node [cate, right of=n1, node distance=5cm, yshift=3cm] (n21) {\textbf{Influence of Friends}};
    \node [cate, below=4.5cm of n21.west, anchor=west] (n22) {\textbf{Preference Integration}};
    \foreach \x in {1,2}
    {\draw[] (n1.east) -- (n2\x.west);}
    
    \node [cate, right of=n21, node distance=8cm, yshift=1.2cm, align=left] (n211) {\textbf{Equal Influence:}
    DiffNet~\cite{wu2019neural}};
    \node [cate, below=1.2cm of n211.west, anchor=west, align=left] (n212) {\textbf{Differentiating Influence by Attention:}\\
    GraphRec~\cite{fan2019graph}, DiffNet++~\cite{wu2020diffnet++}, DANSER~\cite{wu2019dual},\\
    GAT-NSR~\cite{GAT-NSR}, DGRec~\cite{song2019session}, TGRec~\cite{TGRec}};
    \node [cate, below=1.3cm of n212.west, anchor=west, align=left] (n213) {\textbf{Modifying the Relations:}\\
    ESRF~\cite{yu2020enhance}, DiffNetLG~\cite{DiffNetLG}};
    \foreach \x in {1,2,3}
    {\draw[] (n21.east) -- (n21\x.west);}
    
     \node [cate, below=2cm of n213.west, xshift=-1cm, align=left] (n221) {
     \textbf{Separate Graphs}};
     \node [cate, below=2cm of n221.west, anchor=west, align=left] (n222) {\textbf{A Unified Graph}};
     \foreach \x in {1,2}
     {\draw[] (n22.east) -- (n22\x.west);}
     
     \node [cate, below=1.0cm of n213.west, xshift=3.5cm, align=left] (n2211) {
     \textbf{Linearity Combination}: \\
     Equal weights: DiffNet~\cite{wu2019neural}\\
     Different weights: DANSER~\cite{wu2019dual}};
     \node [cate, below=1.4cm of n2211.west, anchor=west, align=left] (n2212) {\textbf{Non-linearity Combination}:\\
     GraphRec~\cite{fan2019graph}, GNN-SoR~\cite{GNN-SoR}, SR-HGNN~\cite{SR-HGNN}};
     \foreach \x in {1,2}
     {\draw[] (n221.east) -- (n221\x.west);}
     
     \node [cate, below=1.25cm of n2212.west, anchor=west, align=left] (n2221) {
     \textbf{A Hierarchical Aggregation Schema}: \\
     DiffNet++~\cite{wu2020diffnet++}, SEFrame~\cite{SEFrame}};
     \draw[] (n222.east) -- (n2221.west);
     
    \end{tikzpicture}}
    \caption{\blue{Summary of representative works in social recommendation.}}
    \label{fig:social-summ-fig}
\end{figure}
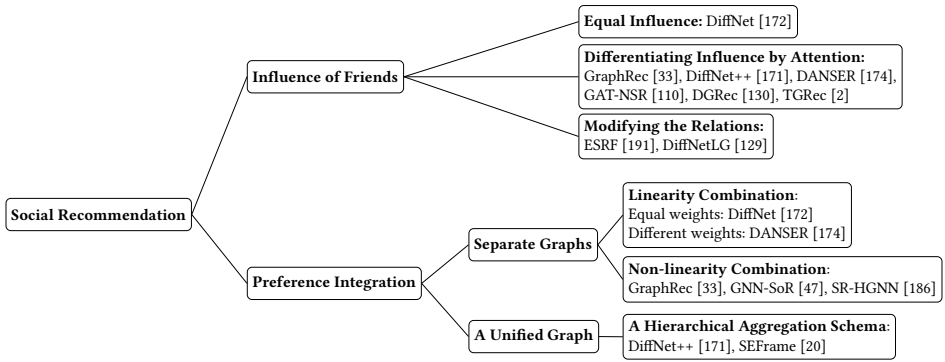
\subsection{Summary}
\blue{Corresponding to the discussion at the beginning of this section, we briefly summarize the current works in terms of the two issues:}

    $\bullet$ \textbf{Influence of Friends.} 
    Compared to assigning equal weights to friends, differentiating the influence of different friends is more appropriate. 
    An emerging direction is to automatically modify the social relationship, which can benefit from the presence of noise in social networks.
    
    $\bullet$ \textbf{Preference Integration.}
    The strategies for combining the two sources of information depends on whether to consider the two graphs separately or unify them into one graph.
    For the separate graphs, user preference is an integration of the overall representations learned from these two graphs.
    For the unified graph, a commonly adopted strategy is the hierarchical aggregation schema.

Figure~\ref{fig:social-summ-fig} summarizes the typical strategies for each of the main issues, and lists the representative works accordingly.

\section{Knowledge Graph based Recommendation}

Social network that reflects relationships between users, is utilized to enhance user representation, while knowledge graph that expresses relationships between items through attributes, is leveraged to enhance the item representation.
Incorporating knowledge graph into recommendation can bring two-facet benefits~\cite{wang2018ripplenet}:
(1) the rich semantic \blue{relations} among items in a knowledge graph can help explore their connections and improve the item representation;
(2) knowledge graph connects a user’s historically interacted items and recommended items, which enhances the interpretability of the results~\blue{\cite{HAGERec}}.

Despite the above benefits, utilizing knowledge graph in recommendation is rather challenging due to its complex graph structure, \emph{i.e.}, multi-type entities and multi-type relations.
Previous works preprocess knowledge graph by knowledge graph embedding (KGE) methods to learn the embeddings of entities and relations, such as \cite{zhang2016collaborative,zhang2018learning,wang2019multi,dadoun2019location}
The limitation of commonly-used KGE methods is that they focus on modeling rigorous semantic relatedness with the transition constraint, which are more suitable for the tasks related to graph, such as link prediction rather than recommendation~\cite{wang2019knowledge}.
\blue{Meta-path based methods manually define meta-paths that carry the high-order information and feed them into a predictive model, and thus they require domain knowledge and are rather labor-intensive for complicated knowledge graph~\cite{wang2019kgat,wang2019kgcn}.}

Given the user-item interaction information as well as the knowledge graph, the knowledge graph based recommendation seeks to take full advantage of the rich information in the knowledge graph, which can help to estimate the users' preferences for items by explicitly capturing relatedness between items.
For the effectiveness of knowledge graph based recommendation, there are two main issues to deal with:

    $\bullet$ \textbf{Graph Construction}. %
    \blue{How to effectively integrate the collaborative signals from user-item interactions and the semantic information from the knowledge graph?
    Whether to explicitly incorporate the user nodes into knowledge graph or implicitly use the user nodes to distinguish the importance of different relations?}
    
    $\bullet$ \textbf{Relation-aware Aggregation}. One characteristic of knowledge graph is that it has multiple types of relations between entities.
    How to design a relation-aware aggregation function to aggregate information from linked entities?

\subsection{Graph Construction}
\blue{For the stage of graph construction, one main concern is how to effectively integrate the collaborative signals and knowledge information.}

\blue{One direction is to incorporate the user nodes into the knowledge graph.
For instance, KGAT~\cite{wang2019kgat}, MKGAT~\cite{sun2020multi} and CKAN~\cite{CKAN} combine the user-item bipartite graph and knowledge graph into one unified graph by taking the user nodes as one type of entity and the relation between users and items as ``interaction".
Recent efforts focus on the entities and relations relevant to the user-item pair. Therefore, they construct the subgraph that links the user-item pair with the user's historical interacted items and the related semantics in knowledge graph~\cite{sha2019attentive,feng2020atbrg}.
Based on the assumption that a shorter path between two nodes reflects more reliable connections, AKGE~\cite{sha2019attentive} constructs the subgraph by the following steps: pre-train the embeddings of entities in the knowledge graph by TransR~\cite{lin2017learning}; calculate the pairwise Euclidean distance between two linked entities; keep the $K$ paths with the shortest distance between the target user and item node.
The potential limitation is that the subgraph structure depends on the pre-trained entity embeddings and the definition of distance measurement.
ATBRG~\cite{feng2020atbrg} exhaustively searches the multi-layer entity neighbors for the target item and the items from the user's historical behaviors and restores the paths connecting the user behaviors and the target item by multiple overlapped entities.
In order to emphasize the information-intensive entities, ATBRG further prunes the entities with a single link, which can also help control the scale of the graph.
Although these methods can obtain subgraphs more relevant to the user-item pair, it is quite time-consuming to pre-train the entity embedding or search and prune paths exhaustively.
An effective and efficient subgraph construction strategy is worthy of further investigation.}

\blue{Another direction is to implicitly use the user nodes to distinguish the importance of different relations.
For instance, KGCN~\cite{wang2019kgcn} and KGNN-LS~\cite{wang2019knowledge} take the user nodes as queries to assign weights to different relations.
In terms of graph construction, this line of research emphasizes the users' preferences towards relations instead of the collaborative signal in user-item interactions.}

\subsection{Relation-aware Aggregation} %
To fully capture the semantic information in knowledge graph, both the linked entities (\emph{i.e.}, $e_i, e_j$) and the relations in between (\emph{i.e.}, $r_{e_i, e_j}$) should be taken into consideration during the propagation process.
Besides, from the perspective of recommender systems, the role of users might also have an influence.
Owing to the advantage of GAT in adaptively assigning weights based on the connected nodes, most of the existing works apply the variants of the traditional GAT over the knowledge graph, \emph{i.e.}, the central node is updated by the weighted average of the linked entities, and the weights are assigned according to the score function, denoted as $a(e_i, e_j, r_{e_i, e_j}, u)$.
The key challenge is to design a reasonable and effective score function.

For the works~\cite{wang2019kgat,sun2020multi,feng2020atbrg} that regard the user nodes as one type of entities, the users' preferences are expected to be spilled over to the entities in the knowledge graph during the propagation process since the item nodes would be updated with the information of interacted users and related attributes, then the other entities would contain users' preferences with iterative diffusion.
Therefore, these works do not explicitly model users' interests in relations but differentiate the influence of entities by the connected nodes and their relations.
For instance, inspired by the transition relationship in knowledge graph, KGAT~\cite{wang2019kgat} assigns the weight according to the distance between the linked entities in the relation space,
\begin{equation}
    a(e_i, e_j, r_{e_i, e_j}, u)=\left(\mathbf{W}_{r} \mathbf{e}_{j}\right)^{\top} \tanh \left(\bigl(\mathbf{W}_{r} \mathbf{e}_{i}+\mathbf{e}_{r_{e_i, e_j}}\bigr)\right),
\end{equation}
where $\mathbf{W}_{r}$ is the transformation matrix for the relation, which maps the entity into relation space.
In this way, the closer entities would pass more information to the central node.
These methods are more appropriate for the constructed subgraph containing user nodes, since it is difficult for the users' interests to extend to all the related entities by stacking a limited number of GNN layers.

For the works that do not combine the two sources of graphs, these studies~\cite{wang2019kgcn,wang2019knowledge} explicitly characterize users' interests in relations by assigning weights according to the connecting relation and specific user.
For example, the score function adopted by KGCN~\cite{wang2019kgcn} is the dot product of the user embedding and the relation embedding, \emph{i.e.}, 
\begin{equation}
    a(e_i, e_j, r_{e_i, e_j}, u)=\mathbf{u}^{\top}\mathbf{r_{e_i, e_j}}.
\end{equation}
In this way, the entities whose relations are more consistent with users' interests will spread more information to the central node.

\tikzset{
  FARROW/.style={arrows={-{Latex[length=1.25mm, width=1.mm]}}, }, %
  U/.style = {circle, draw=melon!400, fill=melon, minimum width=1.4em, align=center, inner sep=0, outer sep=0},
  I/.style = {circle, draw=tea_green!400, fill=tea_green, minimum width=1.4em, align=center, inner sep=0, outer sep=0},
  cate/.style = {rectangle, draw, minimum width=8em, minimum height=2em, align=center, rounded corners=3}, %
  cate2/.style = {rectangle, minimum width=2em, minimum height=2em, align=center, rounded corners=3},%
  encoder/.style = {rectangle, fill=Madang!82, minimum width=10em, minimum height=3em, align=center, rounded corners=3},
  v_rep/.style={
       rectangle split,
       rectangle split part align=base,
       rectangle split horizontal=true,
       rectangle split draw splits=true,
       rectangle split parts=5,
       rectangle split part fill={red!30, blue!20, athens_gray!80, matisse, silver},
       draw=gray, %
       very thin,
       minimum height=1em,
       minimum width=2em,
       inner sep=2.5pt,
       text centered,
       text=gray,
       rounded corners=1
       },
}

\begin{figure}
    \centering
    \resizebox{0.9\linewidth}{!}{
    \begin{tikzpicture}
    \node [cate, distance=4cm, yshift=-0.5cm] (n1) at (0, 0) {\textbf{Knowledge Graph Based}\\ \textbf{Recommendation}};
    \node [cate, right of=n1, node distance=5cm, yshift=1.5cm] (n21) {\textbf{Graph Construction}};
    \node [cate, below=2.5cm of n21.west, anchor=west, align=left] (n22) {\textbf{Relation-aware Aggregation}};
    \foreach \x in {1,2}
    {\draw[] (n1.east) -- (n2\x.west);}
    
    \node [cate, right of=n21, node distance=5cm, yshift=1.2cm, align=left] (n211) {\textbf{User Node as Entity}};
    \node [cate, below=2cm of n211.west, anchor=west, align=left] (n212) {\textbf{Only Knowledge Graph}: 
    KGCN~\cite{wang2019kgcn}, KGNN-LS~\cite{wang2019knowledge}};
    \foreach \x in {1,2}
    {\draw[] (n21.east) -- (n21\x.west);}
    
    \node [cate, right of=n211, node distance=6cm, yshift=0.6cm, align=left] (n2111) {\textbf{A Unified Graph}:\\
    KGAT~\cite{wang2019kgat}, MKGAT~\cite{sun2020multi}, CKAN~\cite{CKAN}};
    \node [cate, below=1.2cm of n2111.west, anchor=west, align=left] (n2112) {\textbf{User-item Subgraph}: \\
    AKGE~\cite{sha2019attentive}, ATBRG~\cite{feng2020atbrg}};
    \foreach \x in {1,2}
    {\draw[] (n211.east) -- (n211\x.west);}
    
    \node [cate, right of=n22, node distance=7cm, yshift=0.5cm, align=left] (n221) {\textbf{Only Relation}:
    KGAT~\cite{wang2019kgat}, MKGAT~\cite{sun2020multi}, CKAN~\cite{CKAN}};
    \node [cate, below=1cm of n221.west, anchor=west, align=left] (n222) {\textbf{User-aware}: 
    KGCN~\cite{wang2019kgcn}, KGNN-LS~\cite{wang2019knowledge}, KGPolicy~\cite{KGPolicy}};
    \foreach \x in {1,2}
    {\draw[] (n22.east) -- (n22\x.west);}
    
    \end{tikzpicture}}
    \caption{\blue{Summary of representative works in knowledge graph based recommendation.}}
    \label{fig:kg-summ-fig}
\end{figure}
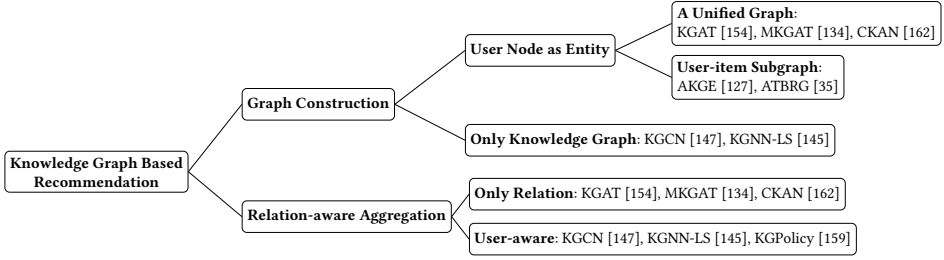
\subsection{Summary}
\blue{Corresponding to the discussion at the beginning of this section, we briefly summarize the current works in terms of the two issues:}

    $\bullet$ \textbf{Graph Construction.} 
    Existing works either consider the user nodes as one type of entities or implicitly use the user nodes to differentiate the relations. 
    The first direction can be further divided into the overall unified graph or the specific subgraph for the user-item pair.
    Compared to the overall unified graph, the user-item subgraph has the advantage of focusing on more related entities and relations, but it requires more computation time and the performance depends on the construction of the subgraph, which still requires further investigation.
    
    $\bullet$ \textbf{Relation-aware Aggregation.}
    The variants of GAT are widely leveraged to aggregate information from linked entities, taking into account the relations.
    For the graphs that do not explicitly incorporate user nodes, user representations are utilized to assign weights to the relations.

Figure~\ref{fig:kg-summ-fig} summarizes the typical strategies for each of the main issues, and lists the representative works accordingly.

\section{Other tasks}
\blue{In addition to these four types of tasks, the researchers have started to utilize GNN for improving the performance of other recommendation tasks, such as POI recommendation and multimedia recommendation.}
In this section, we will summarize the recent developments for each task respectively.

\textbf{Points-of-interest (POI) recommendation} plays a key role in location-based service, which utilizes the geographical information to capture geographical influence among POIs and users' historical check-ins to model the transition patterns.
In the field of POI recommendation, there are several kinds of graph data, such as the user-POI bipartite graph, the sequence graph based on check-ins and geographical graph, \emph{i.e.}, the POIs within a certain distance are connected and the edge weights depend on the distance between POIs~\cite{chang2020learning,lim2020stp}.
\blue{SGRec~\cite{li2021discovering} enriches the check-in sequence with the correlated POIs belonging to other check-ins, which allows collaborative signals to be propagated across sequences.}
\citet{chang2020learning} believe that the more often users consecutively visited the two POIs, the greater the geographical influence between these two POIs.
Hence, the check-ins not only reflect users' dynamic preferences but also indicate the geographical influence among POIs.
To explicitly incorporate the information of geographical distribution among POIs, the edge weights in the sequence graph depend on the distance between POIs~\cite{chang2020learning}.

\textbf{Group recommendation} aims to suggest items to a group of users instead of an individual one~\cite{he2020game} based on their historical behaviors.
There exist three types of relationships: user-item, each user interacts with several items; user-group, a group consists of several users; and group-item, a group of users all choose the same item.
``Group'' can be regarded as a bridge connecting the users and the items in the group recommendation, which can be either treated as a part of the graph or not.
Here are two representative works corresponding to these two strategies respectively.
GAME~\cite{he2020game} introduces the ``group node'' in the graph, and applies the GAT to assign appropriate weights to each interacted neighbor.
With the propagation diffusion, group representation can be iteratively updated with interacted items and users.
However, this approach can not be directly applied to the task where groups are changed dynamically and new groups are constantly formed.
Different from the former transductive method, GLS-GRL~\cite{wang2020group} learns the group representative in an inductive way, which constructs the corresponding graph for each group specifically.
The group representation is generated by integrating the user representations involved in the group, which can address the new group problem.

\textbf{Bundle recommendation} aims to recommend a set of items as a whole for a user.
There are three types of relationships: user-item, each user interacts with several items; user-bundle, users choose the bundles; and bundle-item, a bundle consists of several items.
For group recommendation, ``group'' is made up of users; for bundle recommendation, ``group'' means a set of items.
Analogously, the key challenge is to obtain the bundle representation.
BGCN~\cite{chang2020bundle} unifies the three relationships into one graph and designs the item level and bundle level propagation from the users' perspective.
HFGN~\cite{li2020hierarchical} considers the bundle as the bridge that users interact with the items through bundles.
Correspondingly, it constructs a hierarchical structure upon user-bundle interactions and bundle-item mappings and further captures the item-item interactions within a bundle.

\textbf{Click-through rate (CTR) prediction} is an essential task for recommender systems in large-scale industrial applications, which predicts the click rate based on the multi-type features.
The key challenges of CTR are to model feature interactions and capture user interests.
Inspired by the information diffusion process of GNN, a recent work, Fi-GNN~\cite{li2019fi}, employs GNN to capture the high-order interactions among features.
Specifically, it constructs a feature graph, where each node corresponds to a feature field and different fields are connected with each other through edges.
Hence, the task of feature interactions is converted to propagate node information across the graph.
Despite its considerable performance, Fi-GNN ignores the collaborative signals implicit in user behaviors.
\blue{DG-ENN~\cite{guo2021dual} designs both the attribute graph and user-item collaborative graph, and utilizes GNN techniques to capture the high-order feature interactions and collaborative signals.
To further alleviate the sparsity issue of the user-item interactions, DG-ENN enriches the original user-item interaction relationships with user-user similarity relationships and item-item transitions.}

\blue{\textbf{Multimedia Recommendation} has been a core service to help users identify multimedia contents of interest.
The main characteristic is the contents are in multi-modality, e.g., text, images, and videos.
Recently, researchers start to adopt GNN to capture the collaborative signals from users' interactions with multi-modal contents.
For instance, MMGCN~\cite{wei2019mmgcn} constructs a user-item bipartite graph for each modality and applies GNN to propagate information for each graph, respectively.
The overall user/item representations are the sum of the user/item representations of different modalities.
GRCN~\cite{wei2020graph} utilizes the multi-modal contents to refine the connectivity of user-item interactions.
For each propagation layer, GRCN takes the maximum value of the user-item similarities in different modalities as the weight of the user-item interaction edges and uses the corresponding weights to aggregate neighbors.
MKGAT~\cite{sun2020multi} unifies the user nodes and multi-modal knowledge graph into one graph and employs a relation-aware graph attention network to propagate information.
Considering the multi-modal characteristic of entities, MKGAT designs the entity encoder to map each specific data type into a condensed vector.}

\section{Datasets, Evaluation Metrics and Applications}
\blue{In this section, we introduce the commonly-adopted datasets and evaluation metrics for different recommendation tasks, and summarize the real-world applications of GNN-based recommendation.
This section can help researchers find suitable datasets and evaluation metrics to test their methods, and get an overview of the practical applications of GNN-based recommendation.} %

\subsection{Datasets}
\begin{table*}
    \caption{\blue{Datasets of GNN-based recommendation tasks.}}
    \centering
    \resizebox{0.98\linewidth}{!} {
    \begin{tabular}{l|l|l}
    \toprule
    \textbf{Task} & \textbf{Dataset} & \textbf{Paper}\\
    \midrule
    \multirow{6}{*}{\makecell[l]{User-Item CF}} & MovieLens-100K & GC-MC~\cite{van2018graph}, STAR-GCN~\cite{zhang2019star}, DHCF~\cite{ji2020dual}\\
    & MovieLens-1M & GC-MC~\cite{van2018graph}, SpectralCF~\cite{zheng2018spectral}, STAR-GCN~\cite{zhang2019star}, Bi-HGNN~\cite{li2019hierarchical}, DGCN-BinCF~\cite{wang2019binarized}\\
    & MovieLens-10M & GC-MC~\cite{van2018graph}, STAR-GCN~\cite{zhang2019star}, DGCN-BinCF~\cite{wang2019binarized}\\
    & Amazon & SpectralCF~\cite{zheng2018spectral}, NGCF~\cite{wang2019neural}, LR-GCCF~\cite{chen2020revisiting}, LightGCN~\cite{he2020lightgcn}, DGCF~\cite{wang2020disentangled}, AGCN~\cite{wu2020joint}, NIA-GCN~\cite{sun2020neighbor}, Multi-GCCF~\cite{sun2019multi}\\
    & Gowalla & NGCF~\cite{wang2019neural}, LR-GCCF~\cite{chen2020revisiting}, LightGCN~\cite{he2020lightgcn}, HashGNN~\cite{tan2020learning}, DGCF~\cite{wang2020disentangled}, NIA-GCN~\cite{sun2020neighbor}, Multi-GCCF~\cite{sun2019multi}\\
    & Yelp & NGCF~\cite{wang2019neural}, LightGCN~\cite{he2020lightgcn}, DGCN-BinCF~\cite{wang2019binarized}, DGCF~\cite{wang2020disentangled}, Multi-GCCF~\cite{sun2019multi}\\
    \midrule
    \multirow{7}{*}{\makecell[l]{Sequential\\Recommendation}} & Yoochoose & SR-GNN~\cite{wu2019session}, NISER~\cite{gupta2019niser}, FGNN~\cite{qiu2019rethinking}, HetGNN~\cite{wang2020beyond}, DGTN~\cite{zheng2020dgtn}, DAT-MDI\cite{chen2021dual}, SGNN-HN~\cite{pan2020star}, SHARE~\cite{wang2021session}, DHCN~\cite{xia2021self}, \\
    & \multirow{2}{*}{\makecell[l]{Diginetica}} & SR-GNN~\cite{wu2019session}, GC-SAN~\cite{xu2019graph}, NISER~\cite{gupta2019niser}, FGNN~\cite{qiu2019rethinking}, DGTN~\cite{zheng2020dgtn}, GCE-GNN~\cite{wang2020global}, DAT-MDI\cite{chen2021dual},\\
    &  & TASRec~\cite{zhou2021temporal}, COTREC~\cite{xia2021supervised}, SGNN-HN~\cite{pan2020star}, LESSR~\cite{chen2020handling}, SHARE~\cite{wang2021session}, DHCN~\cite{xia2021self} \\
    & Retailrocket & GC-SAN~\cite{xu2019graph}, NISER~\cite{gupta2019niser}, TASRec~\cite{zhou2021temporal}, COTREC~\cite{xia2021supervised} \\
    & Amazon  & MA-GNN~\cite{ma2019memory}, TGSRec~\cite{fan2021continuous} \\
    & LastFM  & GAG~\cite{qiu2020gag}, LESSR~\cite{chen2020handling} \\
    & Gowalla  & GAG~\cite{qiu2020gag}, LESSR~\cite{chen2020handling}, SERec~\cite{SEFrame}, DGSR~\cite{zhang2021dynamic}\\
    \midrule
    \multirow{6}{*}{\makecell[l]{Social\\Recommendation}} & Yelp & DiffNet~\cite{wu2019neural}, DiffNet++~\cite{wu2020diffnet++}, DGRec~\cite{song2019session}, GNN-SoR~\cite{GNN-SoR}, KCGN~\cite{KCGN}\\
    & Flickr & DiffNet~\cite{wu2019neural}, DiffNet++~\cite{wu2020diffnet++}, MHCN~\cite{MHCN}\\
    & Ciao & GraphRec~\cite{fan2019graph}, GNN-SoR~\cite{GNN-SoR}, SR-HGNN~\cite{SR-HGNN}\\
    & Epinions & GraphRec~\cite{fan2019graph}, DANSER~\cite{wu2019dual}, KCGN~\cite{KCGN}, SR-HGNN~\cite{SR-HGNN}\\
    & LastFM & ESRF~\cite{yu2020enhance}, GNN-SoR~\cite{GNN-SoR}, MHCN~\cite{MHCN}\\
    & Douban & ESRF~\cite{yu2020enhance}, DGRec~\cite{song2019session}, SR-HGNN~\cite{SR-HGNN}, MHCN~\cite{MHCN}\\
    \midrule
    \multirow{9}{*}{\makecell[l]{KG based\\Recommendation}} & MovieLens-1M & AKGE~\cite{sha2019attentive}, TGCN~\cite{chen2020tgcn}, KHGT~\cite{KHGT}\\
    & MovieLens-10M  & MKGAT~\cite{sun2020multi} \\
    & MovieLens-20M  & KGCN~\cite{wang2019kgcn}, KGNN-LS~\cite{wang2019knowledge}, CKAN~\cite{CKAN}\\
    & Book-Crossing  & KGCN~\cite{wang2019kgcn}, KGNN-LS~\cite{wang2019knowledge}, CKAN~\cite{CKAN}\\
    & LastFM & KGCN~\cite{wang2019kgcn}, KGNN-LS~\cite{wang2019knowledge}, KGAT~\cite{wang2019kgat}, AKGE~\cite{sha2019attentive}, TGCN~\cite{chen2020tgcn}, CKAN~\cite{CKAN}\\
    & Dianping & KGNN-LS~\cite{wang2019knowledge}, MKGAT~\cite{sun2020multi}, CKAN~\cite{CKAN}\\
    & Yelp & KGAT~\cite{wang2019kgat}, AKGE~\cite{sha2019attentive}, KHGT~\cite{KHGT}\\
    & Amazon & KGAT~\cite{wang2019kgat}, IntentGC~\cite{zhao2019intentgc}, ATBRG~\cite{feng2020atbrg}\\
    & Taobao & IntentGC~\cite{zhao2019intentgc}, ATBRG~\cite{feng2020atbrg}\\
    \bottomrule
    \end{tabular}
    }
    \label{tab:dataset}
\end{table*}

\blue{This part introduces the public commonly-adopted datasets for GNN-based recommendation systems, as summarized in Table~\ref{tab:dataset}.
Due to the page limit, we do not list the datasets used by other recommender tasks, and we refer readers to the published works.}

\blue{\noindent\textbf{MovieLens}\footnote{https://grouplens.org/datasets/movielens/} datasets are collected from MovieLens website, among which three stable benchmark datasets with different scales of rating pairs, \emph{i.e.}, MovieLens-100K, MovieLens-1M, and MovieLens-20M, are most commonly used~\cite{movielens}.
Each dataset contains user-item rating pairs with timestamps, the movie’s attributes and tags, and user demographic features.
The ratings range from 1 to 5, with a minimum interval of 1.
The MovieLens datasets are widely adopted as benchmark datasets in user-item collaborative filtering tasks and knowledge graph based recommendation.}

\blue{\noindent\textbf{Amazon}\footnote{http://jmcauley.ucsd.edu/data/amazon/links.html} dataset includes reviews (ratings, text, helpfulness votes), product metadata (descriptions, category information, price, brand, and image features), and links (also viewed/also bought graphs)~\cite{amazon}.
The full dataset is split into sub-datasets by categories, e.g., Amazon-Books, Amazon-Instant Video, and Amazon-Electronics.
The sub-datasets in Amazon are usually adopted to test the performance in user-item collaborative filtering and sequential recommendation.}

\blue{\noindent\textbf{Yelp}~\footnote{https://www.yelp.com/dataset} dataset contains the user check-ins and is still being updated.
Yelp dataset is widely adopted in user-item collaborative filtering and POI recommendation tasks.
Existing works usually select one-year data for experiments, e.g., NGCF~\cite{wang2019neural} uses the 2018 edition of the Yelp dataset.}

\blue{\noindent\textbf{Gowalla}\footnote{http://snap.stanford.edu/data/loc-gowalla.html} dataset is the check-in dataset obtained from Gowalla, where users share their locations by checking-in~\cite{gowalla}.
In addition to the check-in information, Gowalla dataset also contains the social relationship among users.
Gowalla is a classical dataset for POI recommendation and adopted in user-item collaborative filtering and sequential recommendation as well.}

\blue{\noindent\textbf{Yoochoose}\footnote{http://2015.recsyschallenge.com/challege.html} dataset is obtained from the RecSys Challenge 2015, which contains a stream of user clicks on an e-commerce website within 6 months.
Instead of the entire dataset, most of the recent studies use the most recent fractions 1/64 and 1/4 of the sequences as the experimental datasets, named Yoochoose1/64 and Yoochoose1/4, respectively.}

\blue{\noindent\textbf{Diginetica}\footnote{http://cikm2016.cs.iupui.edu/cikm-cup} is provided by CIKM Cup 2016, which contains the transactional data in chronological order.
Diginetica is commonly used in session-based recommendation.}

\blue{\noindent\textbf{RetailRocket}\footnote{https://www.kaggle.com/retailrocket/ecommerce-dataset} dataset has been collected from a real-world ecommerce website, which contains six months of user browsing activities.}

\blue{\noindent\textbf{LastFM\footnote{http://mtg.upf.edu/static/datasets/last.fm/lastfm-dataset-1K.tar.gz}} dataset contains musician listening information from a set of 2 thousand users and the attributes of artists from Last.fm\footnote{https://www.last.fm/} online music system~\cite{lastfm}.
This dataset is widely adopted by sequential recommendation, social recommendation and knowledge graph based recommendation.}

\blue{\noindent\textbf{Epinions} dataset and \textbf{Ciao} dataset are shared by \citet{tang2012etrust}.
Each dataset contains the users' ratings (from 1 to 5), reviews towards items, and the directed trust relationships between users.
These two datasets have been widely used as benchmarks for social recommendation.}

\blue{\noindent\textbf{Book-Crossing}\footnote{http://www2.informatik.uni-freiburg.de/~cziegler/BX/} dataset contains 1 million ratings (ranging from 0 to 10) of books and the attributes of books (e.g., title, author) in the Book-Crossing community.
This dataset is widely used as a benchmark for knowledge graph based recommendation.}

\subsection{Evaluation Metrics}
\begin{table*}
    \caption{\blue{Evaluation metrics of GNN-based recommendation tasks.}}
    \centering
    \resizebox{\linewidth}{!} {
    \begin{tabular}{l|l|l}
    \toprule
    \textbf{Task} & \textbf{Metric} & \textbf{Paper}\\
    \midrule
    \multirow{4}{*}{\makecell[l]{User-Item CF}} & Recall & SpectralCF~\cite{zheng2018spectral}, NGCF~\cite{wang2019neural}, LightGCN~\cite{he2020lightgcn}, DGCN-BinCF~\cite{wang2019binarized}, DHCF~\cite{ji2020dual}, DGCF~\cite{wang2020disentangled}, NIA-GCN~\cite{sun2020neighbor}, Multi-GCCF~\cite{sun2019multi}\\
    & MAP & SpectralCF~\cite{zheng2018spectral}, DGCN-BinCF~\cite{wang2019binarized}\\
    & NDCG & NGCF~\cite{wang2019neural}, LR-GCCF~\cite{chen2020revisiting}, LightGCN~\cite{he2020lightgcn}, AGCN~\cite{wu2020joint}, HashGNN~\cite{tan2020learning}, DGCN-BinCF~\cite{wang2019binarized}, DHCF~\cite{ji2020dual}, DGCF~\cite{wang2020disentangled}, NIA-GCN~\cite{sun2020neighbor}, Multi-GCCF~\cite{sun2019multi}\\
    & HR & LR-GCCF~\cite{chen2020revisiting}, AGCN~\cite{wu2020joint}, HashGNN~\cite{tan2020learning}, DHCF~\cite{ji2020dual}, PinSage~\cite{ying2018graph}\\
    \midrule
    \multirow{6}{*}{\makecell[l]{Sequential\\Recommendation}} & Precision & SR-GNN~\cite{wu2019session}, FGNN~\cite{qiu2019rethinking}, GCE-GNN~\cite{wang2020global}, COTREC~\cite{xia2021supervised}, SGNN-HN~\cite{pan2020star}, DHCN~\cite{xia2021self}\\
    & \multirow{2}{*}{\makecell[l]{MRR}} & SR-GNN~\cite{wu2019session}, GC-SAN~\cite{xu2019graph}, NISER~\cite{gupta2019niser}, FGNN~\cite{qiu2019rethinking}, GAG~\cite{qiu2020gag}, HetGNN~\cite{wang2020beyond}, A-PGNN~\cite{wu2019personalizing}, DGTN~\cite{zheng2020dgtn}, GCE-GNN~\cite{wang2020global},\\ 
    & & COTREC~\cite{xia2021supervised}, SGNN-HN~\cite{pan2020star}, LESSR~\cite{chen2020handling}, SURGE~\cite{chang2021sequential}, SHARE~\cite{wang2021session}, DHCN~\cite{xia2021self}, SERec~\cite{SEFrame}, TGSRec~\cite{fan2021continuous}\\
    & HR & GC-SAN~\cite{xu2019graph}, HetGNN~\cite{wang2020beyond}, LESSR~\cite{chen2020handling}, SHARE~\cite{wang2021session}, SERec~\cite{SEFrame}, DGSR~\cite{zhang2021dynamic}\\
    & Recall & NISER~\cite{gupta2019niser}, GAG~\cite{qiu2020gag}, A-PGNN~\cite{wu2019personalizing}, DGTN~\cite{zheng2020dgtn}, DAT-MDI\cite{chen2021dual}, TASRec~\cite{zhou2021temporal}, MA-GNN~\cite{ma2019memory}, TGSRec~\cite{fan2021continuous}\\
    & NDCG & HetGNN~\cite{wang2020beyond}, DAT-MDI\cite{chen2021dual}, TASRec~\cite{zhou2021temporal}, MA-GNN~\cite{ma2019memory}, SURGE~\cite{chang2021sequential}, DGSR~\cite{zhang2021dynamic}\\
    \midrule
    \multirow{5}{*}{\makecell[l]{Social\\Recommendation}} & HR &  DiffNet~\cite{wu2019neural}, DiffNet++~\cite{wu2020diffnet++}, KCGN~\cite{KCGN}\\
    & NDCG & DiffNet~\cite{wu2019neural}, DiffNet++~\cite{wu2020diffnet++}, ESRF~\cite{yu2020enhance}, DGRec~\cite{song2019session}, GNN-SoR~\cite{GNN-SoR}, KCGN~\cite{KCGN}, MHCN~\cite{MHCN}\\
    & AUC & DANSER~\cite{wu2019dual}, HGP~\cite{HGP}\\
    & Precision & DANSER~\cite{wu2019dual}, ESRF~\cite{yu2020enhance}, MHCN~\cite{MHCN}\\
    & Recall & ESRF~\cite{yu2020enhance}, DGRec~\cite{song2019session}, MHCN~\cite{MHCN}\\
    \midrule
    \multirow{5}{*}{\makecell[l]{KG based\\Recommendation}} & AUC & KGCN~\cite{wang2019kgcn}, IntentGC~\cite{zhao2019intentgc}, ATBRG~\cite{feng2020atbrg}, CKAN~\cite{CKAN}\\
    & F1 & KGCN~\cite{wang2019kgcn}, CKAN~\cite{CKAN}\\
    & Recall & KGNN-LS~\cite{wang2019knowledge}, KGAT~\cite{wang2019kgat}, MKGAT~\cite{sun2020multi}\\
    & NDCG & KGAT~\cite{wang2019kgat}, AKGE~\cite{sha2019attentive}, MKGAT~\cite{sun2020multi}, TGCN~\cite{chen2020tgcn}, KHGT~\cite{KHGT}\\
    & HR & AKGE~\cite{sha2019attentive}, TGCN~\cite{chen2020tgcn}, KHGT~\cite{KHGT}\\
    \bottomrule
    \end{tabular}
    }
    \label{tab:eval}
\end{table*}

\blue{It is essential to select adequate metrics to evaluate the performance of compared methods.
Table~\ref{tab:eval} summarizes the evaluation metrics adopted by different recommendation tasks.}

\blue{\noindent\textbf{HR} measures the proportion of users who have at least one click on the recommended items, \emph{i.e.,}
\begin{equation}
    \textbf{HR}@K = \frac{1}{|\mathcal{U}|}\Sigma_{u\in \mathcal{U}}I(|R^K(u)\cap T(u)|>0),
\end{equation}
where $T(u)$ denotes the ground truth item set, $R^K(u)$ denotes the top-K recommended item set, and $I(\cdot)$ is the indicator function.}

\blue{\noindent\textbf{Precision}, \textbf{Recall} and \textbf{F1} are widely adopted to evaluate the accuracy of top-K recommendation.
Precision@K measures the fraction of the items the user will click among the recommended K items.
Recall@K measures the proportion of the number of user clicks in the recommended K items to the entire click set.
F1@K is the combination of Precision@K and Recall@K.
\begin{equation}
    \begin{aligned}
    \textbf{Precision}@K(u) =& \frac{|R^K(u)\cap T(u)|}{K},\quad
    \textbf{Recall}@K(u) = \frac{|R^K(u)\cap T(u)|}{|T(u)|},\\
    \textbf{F1}@K(u) =& \frac{2\times \textbf{Precision}@K(u) \times \textbf{Recall}@K(u)}{\textbf{Precision}@K(u)+\textbf{Recall}@K(u)}.
    \end{aligned}
\end{equation}
The overall metric is the average over all the users', e.g., $\textbf{Precision}@K = \frac{1}{|\mathcal{U}|}\Sigma_{u\in \mathcal{U}} \textbf{Precision}@K(u)$.}

\blue{\noindent\textbf{NDCG} differentiates the contributions of the accurately recommended items based on their ranking positions.
\begin{equation}
    \textbf{NDCG}@K = \frac{1}{|\mathcal{U}|}\Sigma_{u\in \mathcal{U}}\frac{\Sigma_{k=1}^K \frac{I(R^K_k(u)\in T(u))}{\log{(k+1)}}}{\Sigma_{k=1}^K \frac{1}{\log{(k+1)}}},
\end{equation}
where $R^K_k(u)$ denotes the $k^{th}$ item in the recommended list $R^K(u)$.}

\blue{\noindent\textbf{MAP} is a widely adopted ranking metric, which measures the average precision over users,
\begin{equation}
    \textbf{MAP}@K = \frac{1}{|\mathcal{U}|}\Sigma_{u\in \mathcal{U}}\Sigma_{k=1}^K \frac{I(R^K_k(u)\in T(u))\mathbf{Precision}@k(u)}{K}.
\end{equation}}

\blue{\noindent\textbf{AUC} is the probability that the model ranks a clicked item more highly than a non-clicked item.
When the implicit feedback estimation is considered a binary classification problem, AUC is widely adopted to evaluate the performance.
\begin{equation}
    \textbf{AUC}(u) = \frac{\Sigma_{i\in T(u)}\Sigma_{j\in \mathcal{I}\setminus T(u)} I(\hat{r}_i>\hat{r}_j)}{|T(u)||\mathcal{I}\setminus T(u)|}.
\end{equation}
The overall AUC is the average over all the users'.}

\subsection{Applications}
\blue{%
In this part, we summarize the real-world applications of GNN-based recommendation models according to the existing works published by the industry.}

\blue{Product(Advertisement) recommendation on E-commerce platform is one of the most common application scenarios~\cite{li2019hierarchical,tan2020learning,li2020hierarchical,zhao2019intentgc,feng2020atbrg,KHGT}.
For instance, IntentGC~\cite{zhao2019intentgc} leverages both explicit preferences in user-item interactions and heterogeneous relationships in knowledge graph by graph convolutional networks, and is deployed at Alibaba platform for recommending products to users.
Another application is the content recommendation, which recommends news, articles to users.
For example, \citet{wu2019dual} deploy DANSER on a real-world article recommender system, WeChat Top Story by exploiting the user-article interactions and social relationships.
App recommendation has also attempted to utilize GNN-based models, e.g., GraphSAIL is deployed in the recommender system of a mainstream App Store, which selects several hundred apps from the universal set for each user~\cite{xu2020graphsail}.
Besides, \citet{ying2018graph} deploy PinSage at Pinterest, which can generate higher-quality recommendations of images than comparable deep learning and graph-based alternatives.}

\section{Future research directions and open issues}
Whilst GNN has achieved great success in recommender systems, this section outlines several promising prospective research directions.

\subsection{Diverse and Uncertain Representation}

In addition to heterogeneity in data types (e.g., node types like user and item, and edge types like different behavior types), users in the graph usually also have diverse and uncertain interests~\cite{Chen:cikm2020:Improving,Kunaver:kbs2017:diversity}.
Representing each user as a onefold vector (a point in the low-dimensional vector space) as in previous works is hard to capture such characteristics in users' interests.
Thus, how to represent users' multiple and uncertain interests is a direction worth exploring.

A natural choice is to extend such onefold vector to multiple vectors with various methods~\cite{Weston:RecSys13:Nonlinear,Liu:SIGIR20:Octopus,Liu:ijcai2020:Intent}, e.g., disentangled representation learning~\cite{ma2019learning,ma2020disentangled} or capsule networks~\cite{Li:CIKM2019:MIND}.
Some works on GNN-based recommendation also have begun to represent users with multiple vectors.
For instance, DGCF~\cite{wang2020disentangled} explicitly adds orthogonal constraints for multi-aspect representations and iteratively updates the adjacent relationships between the linked nodes for each aspect respectively.
The research of multiple vector representation for recommendation, especially for GNN-based recommendation model, is still in the preliminary stage, and many issues need to be studied in the future, e.g., how to disentangle the embedding pertinent to users' intents; how to set the different interest number for each user in an adaptive way; how to design an efficient and effective propagation schema for multiple vector representations.

Another feasible solution is to represent each user as a density instead of a vector.
Representing data as a density (usually a multi-dimensional Gaussian distribution) provides many advantages, e.g., better encoding uncertainty for a representation and its relationships, and expressing asymmetries more naturally than dot product, cosine similarity, or euclidean distance. 
Specifically, Gaussian embedding has been widely used to model the data uncertainty in various domains, e.g., word embedding~\cite{Vilnis:ICLR2015:Word}, document embedding~\cite{nikolentzos:eacl2017:multivariate,Gourru:ijcai20:Gaussian}, and network/graph embedding~\cite{He:CIKM15:Learning,bojchevski2018deep,Wang:ijcai19:Tag2Gauss}.
For recommendation, \citet{Santos:sigir17:Gaussian} and \citet{Jiang:ijcai19:guassian} also deploy Gaussian embedding to capture users' uncertain preferences for improving user representations and recommendation performance.
Density-based representation, e.g., Gaussian embedding, is an interesting direction that is worth exploring but has not been well studied in the GNN-based recommendation models.

\subsection{Scalability of GNN in Recommendation}
In industrial recommendation scenarios where the datasets include billions of nodes and edges while each node contains millions of features, it is challenging to directly apply the traditional GNN due to the large memory usage and long training time.
To deal with the large-scale graphs, there are two mainstreams: one is to reduce the size of the graph by sampling to make existing GNN applicable; another is to design a scalable and efficient GNN by changing the model architecture.
Sampling is a natural and widely adopted strategy for training large graphs.
For example, GraphSAGE~\cite{hamilton2017inductive} randomly samples a fixed number of neighbors, and PinSage~\cite{ying2018graph} employs the random walk strategy for sampling.
Besides, some works~\cite{sha2019attentive,feng2020atbrg} reconstruct the small-scale subgraph from the original graph for each user-item pair.
However, sampling will lose more or less part of the information, and few studies focus on how to design an effective sampling strategy to balance the effectiveness and scalability.

Another mainstream to solve this problem is to decouple the operations of nonlinearities and collapsing weight matrices between consecutive layers~\cite{he2020lightgcn,sign_icml_grl2020,wu2019simplifying}.
As the neighbor-averaged features need to be precomputed only once, they are more scalable without the communication cost in the model training.  
However, these models are limited by their choice of aggregators and updaters, as compared to traditional GNN with higher flexibility in learning~\cite{chen2020graph}.
Therefore, more future works should be studied in face of the large-scale graphs.

\subsection{Dynamic Graphs in Recommendation}
In real-world recommender systems, not only the objects such as users and items, but also the relationships between them are changing over time. 
To maintain the up-to-date recommendation, the systems should be iteratively updated with the new coming information.
From the perspective of graphs, the constantly updated information brings about dynamic graphs instead of static ones.
Static graphs are stable so they can be modeled feasibly, while dynamic graphs introduce changing structures.
An interesting prospective research problem is how to design the corresponding GNN framework in response to the dynamic graphs in practice.
Existing studies in recommendation pay little attention to the dynamic graphs.
As far as we know, GraphSAIL~\cite{xu2020graphsail} is the first attempt to address the incremental learning on GNN for recommender systems, which deals with the changing of interactions, i.e., the edges between nodes.
To balance the update and preservation, it constraints the embedding similarity between the central node and its neighborhood in successively learned models and controls the incrementally learned embedding close to its previous version.
Dynamic graphs in recommendation is a largely under-explored area, which deserves further studying.

\subsection{Reception Field of GNN in Recommendation}
The reception field of a node refers to a set of nodes including the node itself and its neighbors reachable within $K$-hops~\cite{quan2019brief}, where $K$ is the number of propagation iterations.
\blue{Generally, the aggregation step $K$ is the same as the number of GNN layers in coupled GNNs (e.g., GCN and GraphSAGE). In addition, some recent graph diffusion-based works~\cite{klicpera2019diffusion,klicpera2018predict,miao2021degnn} decouple the operations of aggregation and update and embrace a larger reception field with a larger aggregation step.}
For nodes with low degree, they need deep GNN architecture to enlarge their reception field for sufficient neighborhood information.
However, by increasing the propagation steps, the reception field of nodes with high degree will expand too big and may introduce noise, which could lead to the over-smoothing problem~\cite{li2018deeper} and a consequent drop in performance.

For the graph data in recommendation, the degree of nodes exhibits a long tail distribution, i.e., active users have lots of interactions with items while cold users have few interactions, and similar to the popular items and cold items.
Therefore, applying the same propagation step on all the nodes may be suboptimal.
There are only a few emerging works to adaptively decide the propagation step for each node in order to obtain a reasonable reception field~\cite{liu2020towards,liu2019geniepath,kazi2019inceptiongcn}.
As a result, how to adaptively select a suitable reception field for each user or item in GNN-based recommendation is still an issue worth of research.

\subsection{Self-supervised Learning}
\blue{Self-supervised learning (SSL) is an emerging paradigm for improving the utilization of data, which can help alleviate the sparsity issue.
Inspired by the success of SSL in other areas, recent efforts have leveraged SSL to recommender systems, and made remarkable achievements~\cite{zhou2020s3,wu2021self}.
In the field of GNN-based recommender systems, there exist few attempts to employ SSL as well.
For instance, COTREC~\cite{xia2021supervised} designs a contrastive learning task by maximizing the agreement between the representations of the last-clicked item and the predicted items samples, accompanied with the given session representation.
DHCN~\cite{xia2021self} maximizes the mutual information between the session representations learned via the session-to-session graph and item-session hypergraph.
The key challenge is how to design an effective supervised signal corresponding to the main task.
Considering the prevalence of sparsity issue in recommender systems, we believe self-supervised learning in GNN-based recommender systems is a promising direction.}

\subsection{Robustness in GNN-based Recommendation}
Recent studies show that GNN can be easily fooled by small perturbation on the input~\cite{jin2020adversarial}, i.e., the performance of GNN will be greatly reduced if the graph structure contains noise. 
In real-world recommendation scenarios, it is a common phenomenon that the relationships between nodes are not always reliable.
For instance, users may accidentally click the items, and part of social relationships cannot be captured.
In addition, the attacker may also inject fake data into the recommender systems.
Therefore, it is of great practical significance to construct a robust recommender
system that is able to generate stable recommendations even in the
presence of shilling attacks.
\blue{Due to the vulnerability of GNN to noisy data, it is of great practical significance to construct a robust recommender system that is able to generate stable recommendations even in the presence of shilling attacks~\cite{zhang2020gcn}.
In the field of GNN, there are emerging efforts on graph adversarial learning to enhance the robustness~\cite{he2018adversarial,zhu2019robust,jin2020adversarial}.
Few attempts in GNN-based recommendation have started to pay attention to robustness.
For instance, GraphRf~\cite{zhang2020gcn} jointly learn the rating prediction and fraudster detection, where the probability of being a fraudster determines the reliability of the user's rating in the rating prediction component.
How to build a more robust recommender system is worth exploring but has not been well studied in the GNN-based recommender systems.}

\subsection{Privacy Preserving}
\blue{Due to the strict privacy protection under General Data Protection Regulation~\footnote{https://gdpr-info.eu/}, the privacy preservation in recommender systems has aroused lots of concern in academia and industry since most of the data may be deemed confidential/private, e.g., social network and historical behavior~\cite{he2021stealing}.
An emerging paradigm is to use federated learning to train recommender systems without uploading users' data to the central server
\cite{Muhammad:kdd20:FedFast,Lin:sigir20:Meta,wang:vldbj2021:fast}.
However, the local user data only contains first-order user-item interactions, which is challenging to capture high-order connectivity without privacy linkage~\cite{wu2021fedgnn}.
Another line is to employ differential privacy to guarantee the user privacy in the procedure of recommender systems~\cite{shin2018privacy,Gao:sigir20:DPLCF}.
One limitation of differential privacy is that it usually brings a decrease in performances~\cite{domingo2021limits}.}

\blue{Some efforts have focused on privacy-preserving in GNN-based recommendation. For instance, FedGNN~\cite{wu2021fedgnn} trains the GNN model locally based on the local user-item graph. The pseudo interacted items, and a user-item graph expansion method are proposed to protect the items and exploit the high-order interactions, respectively. Based on local differential privacy (LDP)~\cite{cormode2018privacy}, FedGNN may add noise to the local gradient of each model and thus decrease the model accuracy. To reduce the amount of noise while maintaining privacy protection, PPGRec~\cite{ru2021graph} converts the LDP model into the central differential privacy model and only adds noise to the aggregated global gradient. 
With the increasing society's emphasis on privacy protection, privacy preservation in GNN-based recommendations should be an attractive direction due to its practical value.}

\subsection{Fairness in GNN-based Recommender System}
\blue{Recent years have seen a surge of research effort on recommendation biases to achieve fairness~\cite{chen2020bias}.
For instance, the recommendation performance for users of different demographic groups should be close, and each item should have an equal probability of overall exposure.
With the widespread of GNN, there is an increasing societal concern that GNN could make discriminatory decisions~\cite{dong2021edits}.
Some explorations have been made towards alleviating biases in GNN-based recommender systems.
For instance, NISER~\cite{gupta2019niser} applies normalization operation over the representations to handle popularity bias.
FairGNN~\cite{dai2021say} employs an adversarial learning paradigm to eliminate the bias of GNN by leveraging graph structures and limited sensitive information.
Due to the prevalence of biases in recommender systems and society’s growing focus on fairness, ensuring fairness whilst maintaining comparable performance in GNN-based recommender systems deserves further studying.}

\subsection{Explainability}
\blue{Explainability is beneficial for recommender systems:
on the one hand, explainable recommendations to users allow them to know why the items are recommended and could be persuasive;
on the other hand, the practitioners can know why the model works, which could help further improvements~\cite{zhang2019deep}.
Due to the significance of explainability, many interests have focused on designing explainable recommendation models or conducting post-hoc interpretations~\cite{seo2017interpretable,tay2018multi,Huang:SIGIR2018:Improving}.}

\blue{With the proliferation of GNN, recent efforts have investigated GNN explainability methods~\cite{yuan2020explainability}.
The methods can be divided into two categories:
the instance-level methods provide example-specific explanations by identifying important input features for its prediction~\cite{baldassarre2019explainability,yuan2021explainability,schnake2020higher};
the model-level methods provide high-level interpretations and a generic understanding of how deep graph models work~\cite{yuan2020xgnn}.
There are also some attempts on explainability on GNN-based recommendation~\cite{ma2019jointly,wang2019explainable,huang2019Explainable}.
Most of them utilize semantic information in knowledge graph and conduct post-hoc interpretations.
Up till now, the explainable GNN-based recommender systems are still not fully explored, which should be an interesting and beneficial direction.}

\section{Conclusion}
\blue{Owing to the superiority of GNN in learning on graph data, utilizing GNN techniques in recommender systems has gained increasing interests in academia and industry.}
In this survey, we provided a comprehensive review of the most recent works on GNN-based recommender systems.
We proposed a classification scheme for organizing existing works.
\blue{For each category, we briefly clarified the main issues, detailed the corresponding strategies adopted by the representative models, and discussed their advantages and limitations.}
Furthermore, we suggested several promising directions for future researches.
We hope this survey can provide readers with a general understanding of the recent progress in this field, and shed some light on future developments.

\section*{Acknowledgement}
This work is supported by NSFC (No. 61832001), Beijing Academy of Artificial Intelligence (BAAI), and PKU-Tencent Joint Research Lab.

\bibliographystyle{ACM-Reference-Format}
\bibliography{GNN_RS}
\end{document}